\documentclass[english,aps,preprint,showpacs,preprintnumbers,amsmath,amssymb]{revtex4}
\usepackage[T1]{fontenc}
\usepackage[latin9]{inputenc}
\usepackage{color}
\usepackage{float}
\usepackage{units}
\usepackage{amsmath}
\usepackage{graphicx}
\usepackage{amssymb}

\makeatletter

\providecommand{\tabularnewline}{\\}

\@ifundefined{textcolor}{}
{%
 \definecolor{BLACK}{gray}{0}
 \definecolor{WHITE}{gray}{1}
 \definecolor{RED}{rgb}{1,0,0}
 \definecolor{GREEN}{rgb}{0,1,0}
 \definecolor{BLUE}{rgb}{0,0,1}
 \definecolor{CYAN}{cmyk}{1,0,0,0}
 \definecolor{MAGENTA}{cmyk}{0,1,0,0}
 \definecolor{YELLOW}{cmyk}{0,0,1,0}
 }

\@ifundefined{definecolor}
 {\usepackage{color}}{}
\usepackage{amsfonts}
\usepackage{epsfig}\usepackage{dcolumn}
\usepackage{bm}
\date{\today}

\makeatother

\usepackage{babel}

\makeatother

\usepackage{babel}

\begin{document}

\title{Production of Charged Higgs Bosons in a 3-3-1 Model at the CERN LHC\\
 \bigskip{}
 }

\author{A. Alves$^{1}$, E. Ramirez Barreto$^{2}$, A. G. Dias$^{2}$\medskip{}
 }

\affiliation{$^{1}$Departamento de Ci\^encias Exatas e da Terra, Universidade
Federal de S\~ao Paulo, \\
 Diadema - SP, 09972-270, Brasil\\}

\affiliation{$^{2}$Centro de Ci\^encias Naturais e Humanas, UFABC,\\
 Santo Andr\'e - SP, 09210-170, Brasil \\
 }
\begin{abstract}
We perform a study of the charged Higgs production from an $SU(3)_{C}\otimes SU(3)_{L}\otimes U(1)_{X}$
model with right-handed neutrinos, postulating a custodial symmetry
which reduces the number of free parameters in the scalar potential.
We compute the cross sections for charged scalars for typical and
new production modes. One of the new $SU(3)_{L}$ neutral gauge bosons,
$Z^{\prime}$, affects some production cross sections distinguishing
the model from other standard model extensions like, for example,
the minimal supersymmetric standard model and general two-Higgs doublets
models. The interplay between the Higgs sector of the model and that
$Z^{\prime}$ gauge boson enhances substantially all the production
rates of the lightest charged Higgs boson, $H_{1}^{\pm}$, at hadron
colliders compared to the MSSM. We found that a large portion of the
parameters space can be probed at the LHC running at 14 TeV center-of-mass
energy in the associated $pp\rightarrow W^{\pm}H_{1}^{\mp}+X$ production
channel in the low luminosity run stage of the experiment.
\end{abstract}
\maketitle

\section{Introduction}

The existence of a charged elementary scalar particle implies necessarily
an extended Higgs sector beyond the standard model (SM) and, thus,
an imprint for new physics. Signals of charged elementary scalars
production and effects related with such particles are an important
part of the research at the colliders. This is well justified because
searching for experimental proof of fundamental scalars is a necessary
step to establish which of the many new proposed theories will be
able to explain the experimental facts.

Charged scalars are part of spectrum of several models for physics
beyond the SM. But the investigations realized until now have mainly
been concentrated in the two-Higgs doublet model (2HDM) and the minimal
supersymmetric standard model (MSSM). At present, the most stringent
experimental constraints for the mass of a charged scalar, which we
denote generically as the $H^{\pm}$, comes from LEP-II and from Tevatron
direct searches. Considering 2HDM the experiments yielded the mass
constraint $M_{H^{\pm}}^{2HDM}\geq78.6$ GeV for the charged Higgs
boson, considering decays only into the channels $H^{+}\rightarrow c\overline{s}$
and $\tau^{+}\nu$, \cite{LEP}. For the MSSM, we have the result
from the CDF Collaboration stating that no signal was found in the
mass region $80$ GeV $\leq M_{H^{\pm}}^{susy}\leq$ $160$ GeV for
the charged Higgs boson \cite{TEVA}.

The ATLAS and the CMS Collaborations have made studies about the production
and detection of charged Higgs considering the pair production, and
the associated production with the quark top or with the charged gauge
boson \cite{ATLA}, \cite{CMShiggs}. These studies have shown that
for the LHC energy and luminosities, the charged Higgses could have
a significant potential for discovery. It gives additional motivation
to perform analyses, in the LHC context, of charged scalars predicted
by models different from the 2HDM and the MSSM.

Interesting extensions of the SM are the class of models based on
$SU(3)_{C}\otimes SU(3)_{L}\otimes U(1)_{X}$ group, known as 3-3-1
models \cite{331ppf},\cite{331tp},\cite{331nd},\cite{svs}. These
models have a symmetry breakdown pattern $SU(3)_{C}\otimes SU(3)_{L}\otimes U(1)_{X}$/$SU(3)_{C}\otimes SU(2)_{L}\otimes U(1)_{Y}$
which can be connected to a scalar field condensation at the TeV scale.
The first proposals of these models were constructed considering three
triplets of scalar fields, but a construction with two triplets of
scalar fields is possible as well \cite{ponce2t}. Taking into account
the symmetry breakdown, 8 degrees of freedom from the scalars
fields are incorporated as longitudinal components of the massive
gauge bosons. Therefore, there are more than one physical scalar left
in the particle spectrum of these models, with at least one charged
state. The characteristics of the charged scalars, like electric charge
and couplings, depend on the specific model version. Detailed analysis
of the interactions and production are then crucial in order to validated
or rule out some of these models.

Besides coinciding with the SM at low energies, these models have
several attractive features. Some of them are the following: the anomalies
cancellation occur only when the number of generations are a multiple
of three, and assuming asymptotic freedom in QCD it is concluded that
there must be three generations; the $SU(3)_{L}$ symmetry restricts
the electroweak mixing angle $\theta_{W}$, furnishing a hint for
an explanation of its value, since $\mathrm{sin}\theta_{W}<0.25$
in the models of Refs. \cite{331ppf,331tp}, and $\mathrm{sin}\theta_{W}<0.375$
in the models of Refs. \cite{331nd,svs}; also, quantization of the
electric charge can be explained in the context of these models \cite{piresq}.

These and other features have motivated many studies concerning the
3-3-1 models. For example, new gauge bosons are predicted by these
models and production analysis of such particles has been investigated
in several aspects in Refs. \cite{gbosons} showing, in general, a
great potential for discovering at the LHC and linear colliders. For
other phenomenological issues involving the 3-3-1 models see \cite{Pheno}.

The version in focus here is known as 3-3-1 model with right-handed
neutrinos (3-3-1RHN). It has a new neutral fermionic field, with the
usual charged and neutral fields, for completing each leptonic triplet
representation \cite{svs}, \cite{331nd}. In order to break the symmetry
three scalar triplets are taken into account. In principle, those
triplets may form several invariant operators through multiplication
resulting in a great set of free parameters. In face of it we, additionally,
impose a sort of global custodial symmetry which reduces the number
of free parameters in the scalar potential \cite{331cs}.

Our main goal was to investigate in the 3-3-1RHN the lightest charged
Higgs boson production at hadron colliders. We analyzed several production
channels of phenomenological interest including Higgs pairs, associated
production to $W$ bosons, associated production to top quarks, and
from top quark decay. As a result we find that the cross sections
for all these processes are at least as high as the MSSM analogues.
The presence of the new gauge boson $Z^{\prime}$ of the model affects
some of the production processes that we have studied endowing the
model some distinguishing features compared to other standard model
extensions. In particular, we found that the process $p\, p\rightarrow W^{\pm}H_{1}^{\mp}$+$X$
at the 14 TeV LHC allows, through a reconstruction analysis, a clear
identification of the charged Higgs boson decaying to a top and a
bottom quark.

This paper is organized as follows: In Sec. II, we present briefly
the 3-3-1RHN version and discuss the reduced potential and the
scalar spectrum. The decay channels and the production of the charged
Higgs at the LHC are shown in the Sec. III. In the Sec. IV, we
comment our results involving the decay of the charged scalar and
the gauge boson. Finally, we present our conclusions in Sec. V.

\section{Description of the Model}

In what follows we resume the representation content of the model
and its essential aspects necessary for developing our work.

The lepton triplets are composed by two neutral fields and a charged
lepton

\emph{\begin{eqnarray}
 &  & \Psi_{aL}\equiv\left[\nu_{aL}\,\, e_{aL}\,\, N_{aL}\right]^{T}\sim\left(\mathbf{3,\,}-1/3\right),\label{eq:ltrip}\end{eqnarray}
 }with the right-handed singlets \begin{eqnarray}
e_{aR} & \sim & \left(\mathbf{1,\,}-1\right);\label{eq:lsing}\end{eqnarray}
 where $a=1,2,3$ is the family index; the numbers in parentheses
refer to the transformation properties under $SU(3)_{L}$ and $U(1)_{X}$
(the color quantum number will be omitted), respectively; $N_{aL}$
are new neutral lepton fields. Right-handed neutrinos could be added,
but they are not relevant for the developments here. For the quarks,

\emph{\begin{eqnarray}
 &  & Q_{mL}\equiv\left[d_{mL}\,\, u_{mL}\,\, D_{mL}\right]^{T}\sim\left(\mathbf{3^{*},\,}0\right),\nonumber \\
 &  & Q_{3L}\equiv\left[u_{3L}\,\, d_{3L}\,\, T_{L}\right]^{T}\sim\left(\mathbf{3,\,}1/3\right),\nonumber \\
 &  & u_{\alpha R}\sim\left(\mathbf{3,\,1,\,}2/3\right),\quad d_{\alpha R}\sim\left(\mathbf{1,\,}-1/3\right),\nonumber \\
 &  & T_{R}\sim\left(\mathbf{3,\,1,\,}2/3\right),\quad D_{mR}\sim\left(\mathbf{1,\,}-1/3\right),\label{eq:quarks}\end{eqnarray}
 }where $m=1,2$ and $\alpha=1,2$,3. $D_{m}$ and $T$ are new quark
fields. Anomalies cancellation requires $Q_{mL}$ to be an $SU(3)_{L}$
antitriplet. Scalar fields are such that they form the following
triplets \begin{eqnarray}
 &  & \eta\equiv\left[\eta^{0}\,\,\eta^{-}\,\,\eta^{\prime0}\right]^{T}\sim\left(\mathbf{3,\,}-1/3\right),\nonumber \\
 &  & \rho\equiv\left[\rho^{+}\,\,\rho^{0}\,\,\rho^{\prime+}\right]^{T}\sim\left(\mathbf{3,\,}2/3\right),\nonumber \\
 &  & \chi\equiv\left[\chi^{0}\,\,\chi^{-}\,\,\chi^{\prime0}\right]^{T}\sim\left(\mathbf{3,\,}-1/3\right)\label{eq:etrip}\end{eqnarray}
 In order to restrict the scalar fields self-interactions it is assumed
an approximate global symmetry $SU(3)_{L^{\prime}}\otimes SU(3)_{R^{\prime}}$
from which we define the tritriplet \begin{equation}
\Phi=\left(\eta\:\rho\:\chi\right)\label{eq:ttrip}\end{equation}
 This object transforms under the global symmetry as follows\[
\Phi\rightarrow\Omega_{L^{\prime}}\Phi\Omega_{R^{\prime}}^{\dagger}\]
 and the invariant potential, containing operators up to dimension
four, is\begin{eqnarray}
V\left(\Phi\right) & = & \mu^{2}Tr\left(\Phi^{\dagger}\Phi\right)+\frac{f}{2}\epsilon_{ijk}\epsilon_{lmn}\left(\Phi_{il}\Phi_{jm}\Phi_{kn}+H.c.\right)\nonumber \\
 & + & \lambda_{1}\left[Tr\left(\Phi^{\dagger}\Phi\right)\right]^{2}+\lambda_{2}Tr\left(\Phi^{\dagger}\Phi\right)^{2}\label{poten}\end{eqnarray}
 Both gauge $U(1)_{X}$ and Yukawa couplings break explicitly $SU(3)_{L^{\prime}}\otimes SU(3)_{R^{\prime}}$.
This is the reason we treat it as an approximated global symmetry.
It has the important consequence of allowing us to have different
vacuum expectation values (VEV) for some of the neutral components
in the scalar triplets giving, in this way, a consistent pattern for
breakdown of gauge symmetries. Assuming the vacuum expectation values
$\langle\eta^{0}\rangle=v/\sqrt{2}$, $\langle\rho^{0}\rangle=u/\sqrt{2}$,
and $\langle\chi^{\prime0}\rangle=w/\sqrt{2}$ the constraint equations
for minimum condition of the potential are as follow;

\begin{eqnarray}
\lambda_{1}\left(w^{2}+v_{w}^{2}\right)+\lambda_{2}v^{2}+\frac{\sqrt{2}}{4v}fuw & = & -\mu^{2}-\delta\mu_{1}^{2}\nonumber \\
\lambda_{1}\left(w^{2}+v_{w}^{2}\right)+\lambda_{2}u^{2}+\frac{\sqrt{2}}{4u}fvw & = & -\mu^{2}-\delta\mu_{2}^{2}\nonumber \\
\lambda_{1}\left(w^{2}+v_{w}^{2}\right)+\lambda_{2}w^{2}+\frac{\sqrt{2}}{4w}fuv & = & -\mu^{2}-\delta\mu_{3}^{2}\label{eq:vinc}\end{eqnarray}
 where $v_{w}^{2}=v^{2}+u^{2}$ ($v_{w}=246$ GeV) and $\delta\mu_{i}$
are the loop corrections involving parameters of the explicitly breaking
symmetry terms. Once the scalar triplets have distinct $U(1)_{X}$
charges and couplings with the fermionic fields the right side of
the above Eqs. (\ref{eq:vinc}) are different, so that there can be
a solution for different values of $v$, $u$, and $w$. Now, $\langle\chi^{\prime0}\rangle$
realizes the break $SU(3)_{L}\otimes U(1)_{X}\rightarrow SU(2)_{L}\otimes U(1)_{Y}$
and we assume that $w$ is bigger than the $v$ and $u$. These two
last values are the scales connected to the breakdown to the electromagnetic
factor $SU(2)_{L}\otimes U(1)_{Y}\rightarrow U(1)_{em}$ and, therefore,
directly related to the particle masses we already known in the SM.

We have that 8 from 18 degrees of freedom of scalar triplets turn
into longitudinal polarization for massive gauge bosons $W^{\pm}$,
$Z$, $U^{\pm}$, $V^{0}$, $V^{0\dagger}$, and $Z^{\prime}$ resulting
from symmetry breakdown. Thus, the physical scalar particle spectrum
of the model has three CP even, one CP odd, one neutral complex, and
two single charged scalars composing the remaining 10 degrees of freedom.
We define the triplets neutral components obtaining VEV in terms of
real and imaginary components as

\begin{equation}
\varphi^{0}=\langle\varphi^{0}\rangle+\frac{1}{\sqrt{2}}\left(\xi_{\varphi}+i\zeta_{\varphi}\right),\label{eq:shift}\end{equation}
 and taking into account a further assumption that $f=-4\sqrt{2}\lambda_{1}w$
for simplifying our analyses we get the following mass expressions:
for the three CP even eigenstates $h_{i}^{0}$

\begin{eqnarray}
M_{h_{1}}^{2} & = & 2\lambda_{2}(2x+1)u^{2},\label{eq:mnh1}\\
M_{h_{2}}^{2} & = & 2\lambda_{2}(2xw^{2}+u^{2}),\label{eq:mnh2}\\
M_{h_{3}}^{2} & = & 2\lambda_{2}\left[(x+1)w^{2}+xu^{2}\right];\label{eq:mnh3}\end{eqnarray}
 for the CP odd eigenstate $A^{0}$ \begin{equation}
M_{A^{0}}^{2}=2\lambda_{2}x\left(v_{w}^{2}\frac{w^{2}}{uv}+uv\right);\label{eq:mpseudoe}\end{equation}
 for the complex neutral eigenstate $H^{0}$\begin{equation}
M_{H^{0}}^{2}=\lambda_{2}\left(2x\frac{u}{v}+1\right)(v^{2}+w^{2});\label{eq:mescom}\end{equation}
 and for the two charged eigenstates $H_{1}^{\pm}$ and $H_{2}^{\pm}$\begin{equation}
M_{H_{1}^{\pm}}^{2}=\lambda_{2}\left(2x\frac{w^{2}}{uv}+1\right)v_{w}^{2}\label{eq:mh1}\end{equation}
 \begin{equation}
M_{H_{2}^{\pm}}^{2}=\lambda_{2}\left(2x\frac{v}{u}+1\right)(u^{2}+w^{2})\label{eq:mh2}\end{equation}
 where $x=\lambda_{1}/\lambda_{2}$.

To avoid any conflict caused by a tree level deviation of the $\rho$
parameter we also take a special relation between the VEVs according
to \begin{equation}
u^{2}=\frac{1-2\mathrm{sw^{2}}}{2\mathrm{cw^{2}}}v_{w}^{2}\label{eq:vevrel}\end{equation}
 where $\mathrm{sw^{2}}$ ($\mathrm{cw^{2}}$) stands for $\mathrm{sin}^{2}\theta_{W}$
($\mathrm{cos}^{2}\theta_{W}$ ), with $\theta_{W}$ being the electroweak
mixing angle. The relation above is obtained when we look for VEVs
values which leads to a match with the tree level SM prediction $\rho=1$.
Equation (\ref{eq:vevrel}) is also what come out as a solution for the
VEVs which cancels the mixing between the neutral massive gauge bosons
$Z$ and $Z^{\prime}$. There are two of vacuum configurations realizing
this: one is independent of the VEV $w$ value and given by Eq. (\ref{eq:vevrel}),
the other is for $w\rightarrow\infty$ with $u$ and $v$ taking any
values satisfying $v_{w}^{2}=v^{2}+u^{2}$ (observe that this last
configuration recovers the case of 2HDM where the tree level $\rho$
parameter is the same as in the SM). Also, as a consequence of Eq.
(\ref{eq:vevrel}) all known fermions in the model have the same gauge
vector and axial couplings like in the SM \cite{331cs}. Taking into
account the value $\mathrm{sw^{2}}\approx0.2321$ we have $v\approx198.5$,
and $u\approx145.3$. With this we now take the LEP limit on the CP
even eigenstate $h_{1}^{0}$, which is equivalent to the SM Higgs
boson, as being $M_{h_{1}}\geq114$ GeV in Eq. (\ref{eq:mnh1}) to
constrain the values of the parameters $\lambda_{1}$ and $\lambda_{2}$
according to \begin{equation}
2\lambda_{1}+\lambda_{2}\geq0.31\label{eq:l1l2}\end{equation}

\begin{figure}[H]
 \centering\includegraphics[scale=0.6]{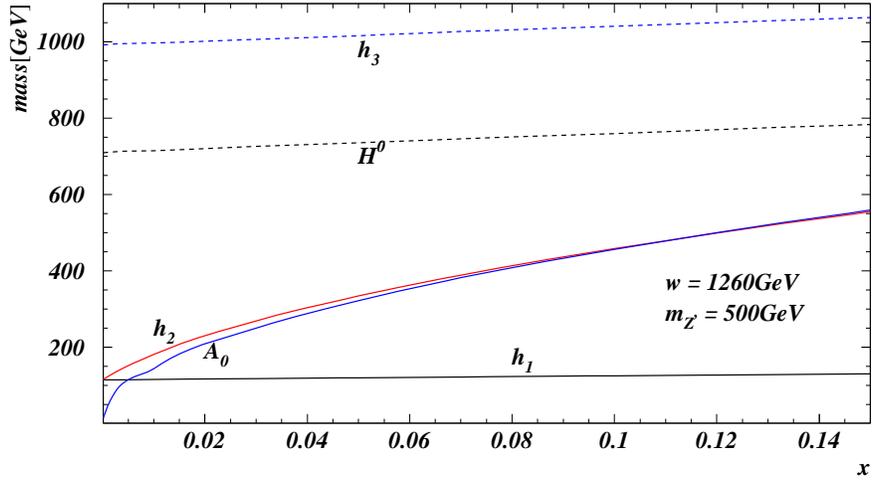} \caption{Masses of the neutral scalar particles $h_{i}^{0}$, $A^{0}$, and
$H^{0}$for $\lambda_{2}=0.31$ and $m_{Z'}=500$ GeV.{\large \label{fig:zl500}}}

\end{figure}

\begin{figure}[H]
 \centering\includegraphics[scale=0.6]{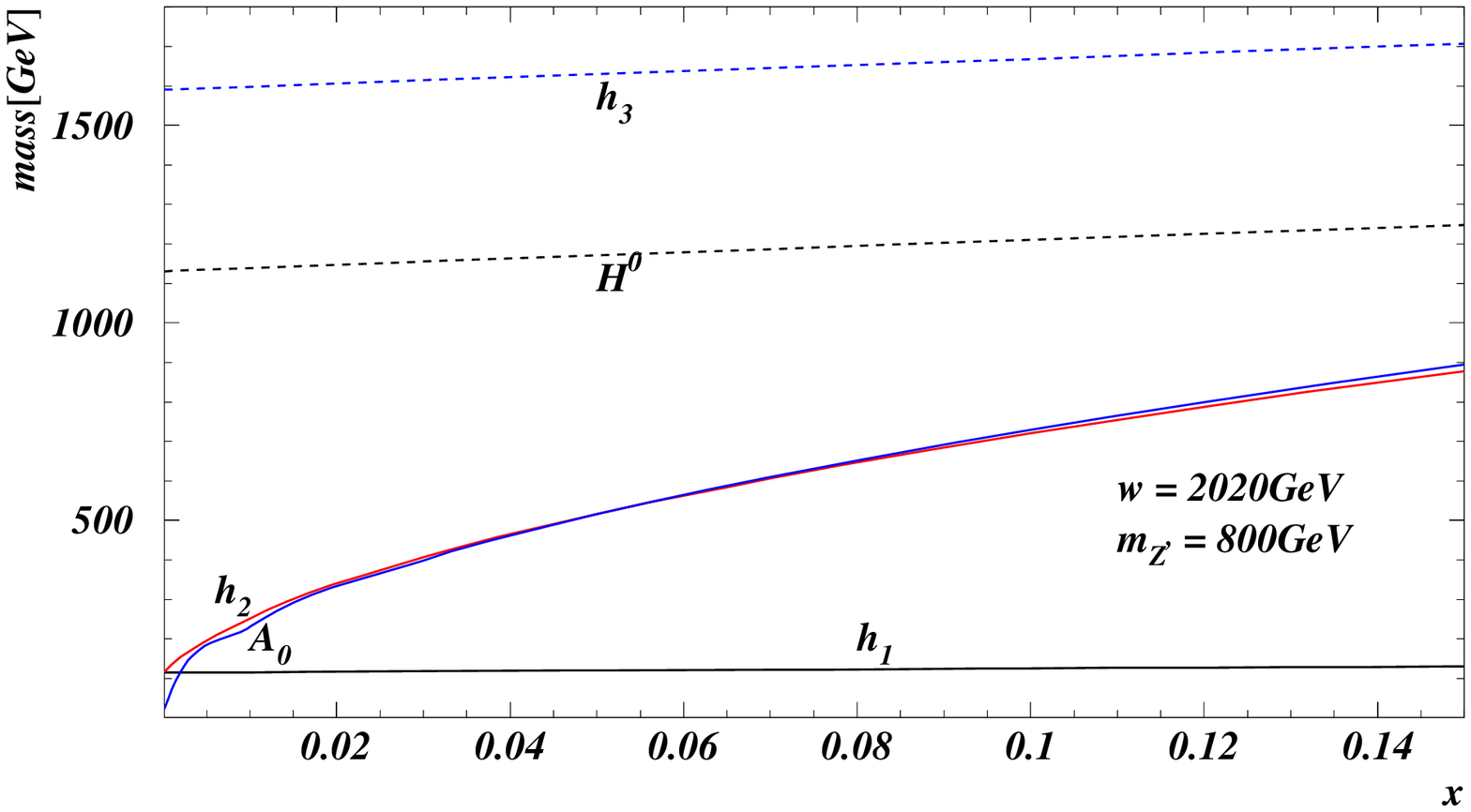} \caption{Masses of the neutral scalar particles $h_{i}^{0}$, $A^{0}$, and
$H^{0}$ for $\lambda_{2}=0.31$ and $m_{Z'}=800$ GeV.{\large \label{fig:zl800}}}

\end{figure}

\begin{figure}[H]
 \centering\includegraphics[scale=0.6]{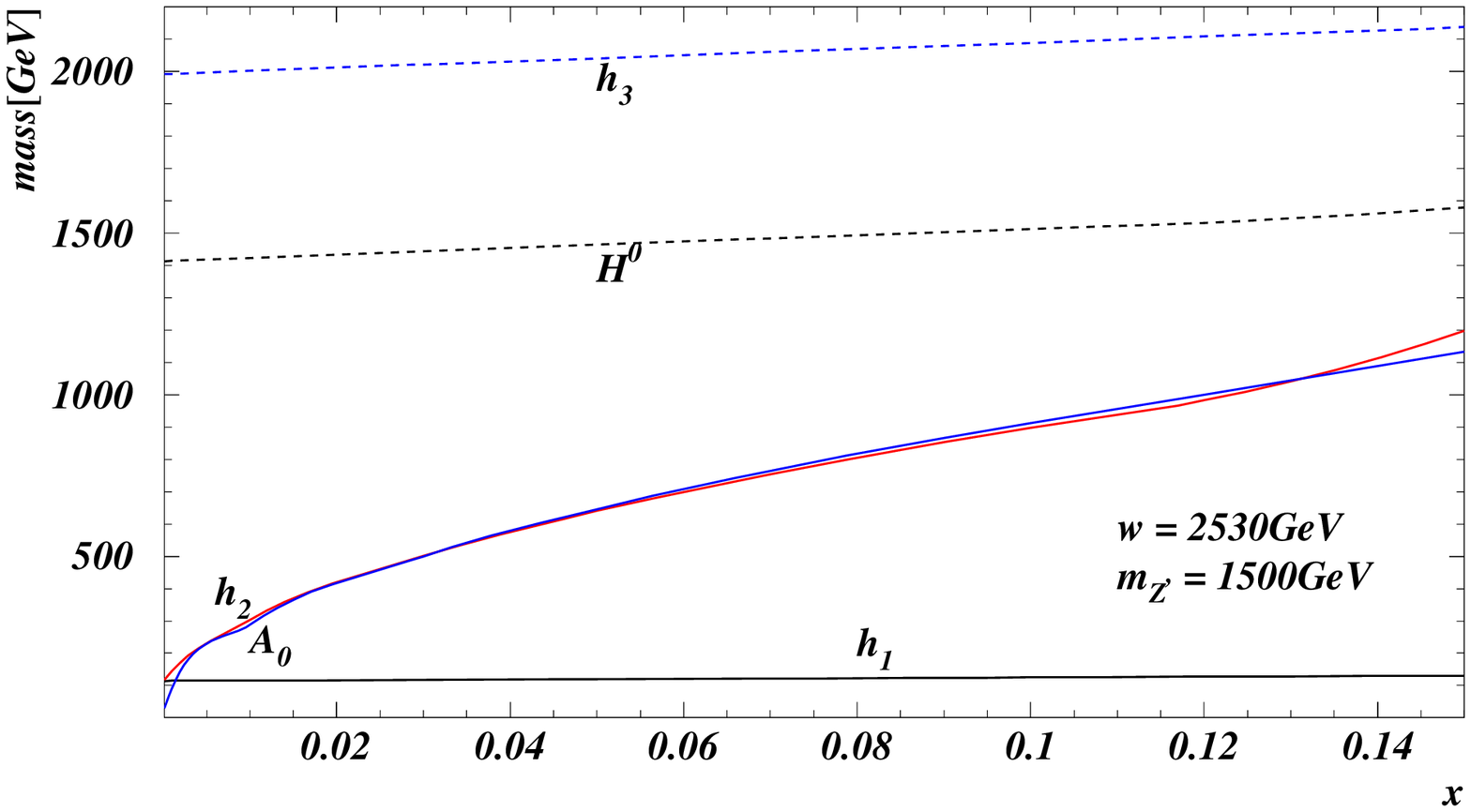} \caption{Masses of the neutral scalar particles $h_{i}^{0}$, $A^{0}$, and
$H^{0}$ for $\lambda_{2}=0.31$ and $m_{Z'}=1500$ GeV.{\large \label{fig:zl1500}}}

\end{figure}

\begin{figure}[H]
 \centering\includegraphics[scale=0.6]{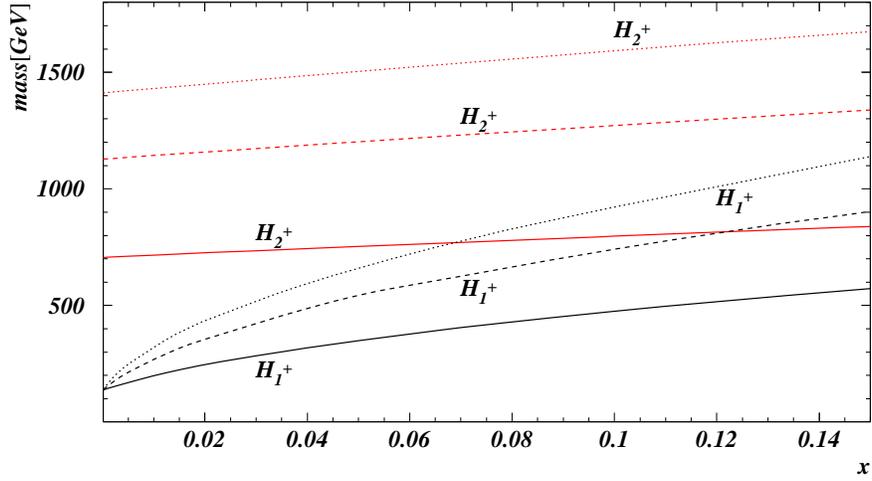} \caption{Masses of the charged scalar particles $H_{1}^{\pm}$ and $H_{2}^{\pm}$
for $\lambda_{2}=0.31$. The solid lines correspond to $m_{Z'}=500$,
the dashed lines to $m_{Z'}=800$ GeV, and the dotted ones to $m_{Z'}=1500$
GeV.{\large {} \label{fig:charged_masses}}}

\end{figure}

Our focus here is on the production of the lightest charged scalar,
and we apply this constraint on the tree level masses obtained above
for the scalar particles. Fixing values for $\lambda_{2}$ and $w$
we show in Figs.  \ref{fig:zl500}, \ref{fig:zl800}, \ref{fig:zl1500}
below the masses of the neutral scalars as a function of $x$ and
in Fig. \ref{fig:charged_masses} the masses of the charged scalars
also as a function of $x$. It is interesting to observe that our
assumption of the approximate global symmetry and the defined values
of $v$ and $u$ make $h_{2}^{0}$, $A^{0}$, and $H_{1}^{\pm}$ practically
degenerated in mass, except for a small interval of values for $x$,
where $A^{0}$ can be lighter than $h_{1}^{0}$. We observe at Fig.
\ref{fig:charged_masses} that the $H_{1}^{\pm}$ are the lighter
ones. Also Eq. (\ref{eq:l1l2}) implies a lower bound of $M_{H_{1}}\geq137$
GeV on the lightest charged scalars. We shall use this lower bound
in our phenomenological analysis of the production and decay of the
lightest charged Higgs bosons.

The gauge bosons $W_{\mu}^{\pm}$ and $Z$ have tree level masses
as in the SM. For the new gauge bosons, $U^{\pm}$, $V^{0}$, $V^{0\dagger}$,
and $Z^{\prime}$, we have the following mass expressions\begin{eqnarray}
m_{U}^{2} & = & \frac{g^{2}}{4}\left(u^{2}+w^{2}\right),\label{eq:Umass}\\
m_{V}^{2} & = & \frac{g^{2}}{4}\left(v^{2}+w^{2}\right),\label{eq:Vmass}\\
m_{Z^{\prime}}^{2} & = & \frac{g^{2}}{36\left(t^{2}+3\right)}\left[\left(2t^{2}+6\right)^{2}w^{2}-\left(8t^{4}-6t^{2}-9\right)v_{w}^{2}\right]\label{eq:zlmass}\end{eqnarray}
 where we have used the definition\begin{equation}
t^{2}=\frac{g_{X}^{2}}{g^{2}}=\frac{\mathrm{sw^{2}}}{1-\frac{4}{3}\mathrm{sw^{2}}}\label{eq:deft}\end{equation}
The tree level mass expression for all new gauge bosons show dependence
with $w$. This is due the fact that such new particles appear when
completing the representation of $SU(3)_{L}\otimes U(1)_{X}$, and
this symmetry is supposedly broken at the $w$ scale.

In the Yukawa sector we assume, just for simplification, that the
known quarks do not have mixing with the new quarks. This can be achieved
taking into account some sort of $Z_{n}$ symmetry which preserves
the scalar potential in Eq. (\ref{poten}). Under this assumption,
the interaction Lagrangian for the known quarks with the charged Higgs
$H_{1}^{\pm}$ is \begin{eqnarray}
-\mathcal{L}_{Y} & = & \frac{g}{\sqrt{2}m_{W}}\frac{v}{u}\overline{\mathcal{U}}M_{u}\left[V_{CKM}-V_{L}^{u\dagger}\Delta V_{L}^{d}\right]P_{L}\mathcal{D}H_{1}^{+}\nonumber \\
 & + & \frac{g}{\sqrt{2}m_{W}}\frac{u}{v}\overline{\mathcal{U}}\left[V_{CKM}-V_{L}^{u\dagger}\Delta V_{L}^{d}\right]M_{d}P_{R}\mathcal{D}H_{1}^{+}+H.c.\label{eq:yuk-h1}\end{eqnarray}
where $\mathcal{D}=(d,s,b)^{T}$, $\mathcal{U}=(u,c,t)^{T}$, with
$V_{CKM}=V_{L}^{u\dagger}V_{L}^{d}$ being the usual Cabbibo-Kobayashi-Maskawa
matrix defined in terms of the rotation matrices, $V_{L}^{u}$ and
$V_{L}^{d}$, for the $u$ and $d$ type quarks mass eigenstates whose
eigenvalues are the entries of the diagonal matrices $M_{u}=diag(m_{u},m_{c},m_{t})$,
$M_{d}=diag(m_{d},m_{s},m_{b})$ and $\Delta=diag(0,0,1)$. The expressions
inside brackets in Eq. (\ref{eq:yuk-h1}) differ from just being $V_{CKM}$,
which is obtained in 2HDMs, for example, because the 3-3-1 model has
a peculiarity that the third generation of quarks transforms diferently
from the other two.

It is clear from the quarks mass matrices that in order to study the
interactions among the charged Higgs and the quarks, we might consider
only contributions of the third generation. Then, we can write the
most relevant interaction in Eq. (\ref{eq:yuk-h1}) involving $H_{1}^{\pm}$
in the form \begin{eqnarray}
-\mathcal{L}_{Y}^{tbH_{1}} & = & \frac{g}{\sqrt{2}m_{W}}\overline{t}\left[V_{CKM}-V_{L}^{u\dagger}\Delta V_{L}^{d}\right]_{33}\left(\frac{v}{u}m_{t}P_{L}+\frac{u}{v}m_{b}P_{R}\right)bH_{1}^{+}+H.c.\label{eq:yuk-h1tb}\end{eqnarray}
 The elements of the matrices $V_{L}^{u}$ and $V_{L}^{d}$ are not
known and we consider here that they are such that $\mid\left[V_{L}^{d\dagger}\Delta V_{L}^{u}\right]_{33}\mid\ll1$.
Therefore, our conclusions will be based on the fact that $H_{1}^{\pm}$
decay mainly through a \emph{top-bottom} channel, when $M_{H_{1}^{\pm}}\geq m_{t}+m_{b}$.

It must be pointed out a similarity of the $H_{1}^{\pm}$ couplings
in Eq. (\ref{eq:yuk-h1tb}) with the corresponding ones in 2HDM. This
is due the fact $H_{1}^{\pm}$ is a combination of symmetry eigenstates
$\eta^{\pm}$ and $\rho^{\pm}$ in Eq. (\ref{eq:etrip}), with each
of these last two fields belonging to two different $SU\left(2\right)_{L}$
doublets. The resemblance turns out more evident putting on Eq. (\ref{eq:yuk-h1tb})
the usual parameter definition $tan\beta=\frac{u}{v}=\sqrt{1-2\mathrm{sw^{2}}}$,
which in our case has a fixed value according to Eq. (\ref{eq:vevrel}).
Thus, in principle, a direct comparison with 2HDM could be done in
these terms. But $H_{1}^{\pm}$ here has new interactions like $H_{1}^{\pm}H_{1}^{\mp}Z'$,
$H_{1}^{\pm}W^{\mp}Z'$ which can enhance its production due $Z'$
contribution in s-channel, for example. We shall discuss more on this
in what follows.

\section{Production of Charged Scalars $H_{1}^{\pm}$ }

The production of lightest charged scalars of the 3-3-1 model of our
study can occur through the following leading modes:
\begin{enumerate}
\item Pair production: $q\overline{q},\, b\overline{b},\, gg\rightarrow H_{1}^{\pm}H_{1}^{\mp}$
\item In association to $W$ bosons: $q\overline{q},\, b\overline{b}\rightarrow H_{1}^{\pm}W^{\mp}$
\item In association to new $Z^{\prime}$ bosons: $q\overline{q'}\rightarrow Z'H_{1}^{\pm}$
\item Single production in association to top quarks: $b\, g\rightarrow tH_{1}^{\pm}$
\item Single production from top quark decay: $q\overline{q},\, gg\rightarrow tbH_{1}^{\pm}$
\end{enumerate}
The production modes (1),(4), and (5) are typical of extended Higgs
sectors as the MSSM for instance. The associated $tH_{1}^{\pm}$ process
(4) initiated by bottom quarks from the proton sea has the largest
production cross section \cite{topHiggs} in the MSSM and general
2HDMs due the $tan\beta$ enhancement. Charged Higgs production from
top quark decays (5) is important for light masses $m_{H_{1}^{\pm}}<m_{t}+m_{b}$,
and its magnitude is comparable to the top pairs production \cite{ttHlight}.
The charged Higgs pair production (1), by its turn, can be substantial
for large $tan\beta$ combining all contributing channels and after
including NLO corrections \cite{alves-plehn}.

The mode (2) can also occur in the MSSM and 2HDM models via t-channel
Yukawa diagrams involving top and bottom quarks, and through s-channel
neutral Higgs bosons \cite{HWmssm}. On the other hand, in the 3-3-1
models, or in models with extended gauge sectors, a new contribution
is possible: through the production and subsequent decay of a new
neutral gauge boson $Z^{\prime}$. In our 3-3-1 model the new interaction
that makes it possible can be read from the expression Eq. (\ref{eq:H1WZl})
in the Appendix. The same interaction is responsible for the mode
(3), the $Z^{\prime}H_{1}^{\pm}$ associated production via an off shell
s-channel $W$ boson. These production modes will give rise to final
state topologies with at least one top quark, $b$ jets, $W$ bosons,
tau leptons, and missing energy. In the next we discuss each production
channel presented above in more detail.

All computations of this work were performed at tree level using the
madgraph/madevent package \cite{Mad} and the signals were checked
against calcHep/compHep \cite{comp}. Higher order corrections were
not taken into account but we comment the expected impact of QCD corrections
wherever we find it is enlightening. The CTEQ6L structure functions
\cite{CETQ} were used for the calculation of all signals and backgrounds.
The factorization scale was chosen as $\mu_{F}=m_{Z^{\prime}}$ for
the simulation of signal events from processes initiated by light
quarks and gluons. The bottom factorization scale was chosen to be
$\nicefrac{\mu_{F}}{4}$ as suggested in \cite{bFAC} and \cite{alves-plehn}
for the sake of stability of the perturbative calculation. Different
choices were made for some specific background processes and will
be discussed in the appropriate moment. The renormalization scale
was set equal to the factorization scale $\mu_{R}=\mu_{F}$ in all
relevant processes. This choice may result in unphysical cancellations
in certain cases but without a higher order analysis the precise impact
of this choice is out of the scope of the study. Anyway, any cancellation
would reduce the production rates which will not spoil our conclusions.

\subsection{Charged Higgs pair production}

Charged Higgs bosons can be produced in pairs through light and bottom
quark annihilation and gluon-gluon fusion. The later channel receives
contributions from loop diagrams with virtual heavy quarks, and the
Higgses are produced via Yukawa interactions to these heavy states.
In principle, all heavy quarks predicted by the theory contribute
to the amplitude, however based on the MSSM case \cite{alves-plehn},
where the gluon fusion channel contributes significantly only for
rather large $tan\beta$, we assume we can neglect that channel for
the computation of total cross sections presenting conservative results.

\begin{figure}[H]
 \centering\includegraphics[scale=0.6]{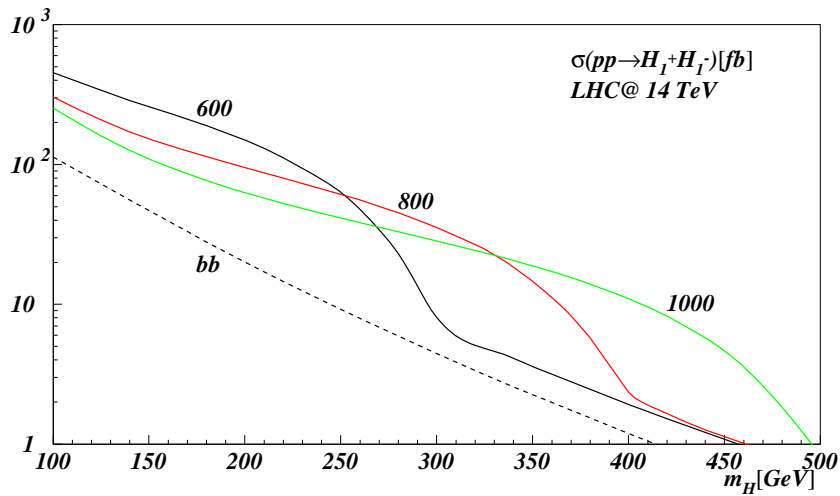} \caption{Production cross section (in fb) of a charged Higgs pair at the 14
TeV LHC including the Drell-Yan contribution and the t-channel bottom
initiated process (solid curves). The dashed curve shows the contribution
from the Yukawa interaction only. We show the total production rates
for three different $Z'$ masses, 600, 800, and 1000 GeV. {\large \label{fig:Drell-Yan}}}

\end{figure}

The Drell-Yan annihilation to neutral gauge bosons and subsequent
decay to Higgs pairs is the main production channel. The production
rate can reach the hundreds of femtobarns level and the new $Z^{\prime}$
contribution is the most important one at the 14 TeV LHC as we can
see in Fig. \ref{fig:Drell-Yan}. As soon as the twice of the Higgs
mass exceeds the $Z^{\prime}$ mass the rates drop sharply and the
bottom initiated process drives the production rates.

The bottom t-channel contribution involves the enhanced Yukawa $tbH_{1}^{\pm}$
interaction
which may compete with the Drell-Yan channel depending on the mass
of the charged Higgs. Nevertheless, the bottom density is populated
for small x in the proton sea, and the production of heavy states is
less likely from this channel.

It is worth noting that the expected rate for the MSSM charged Higgs
pair production including NLO QCD barely reaches 100 fb for $tan\beta=50$,
as demonstrated in Ref.~\cite{alves-plehn} including all possible
contribution channels with little sensitivity to the specific parameter
space point.

The prospects to search for charged Higgses in the pair production
channel was established by the CMS and ATLAS Collaborations \cite{ATLA}
\cite{CMShiggs}, in special the charged scalars predicted by the
2HDM and the MSSM \cite{SUSY1}, \cite{SUSY2} will be hard to detect.
This conclusion can change for left-right symmetric models \cite{lr-2hdm}
and other 2HDM-like models as shown in \cite{logan-2hdm} and where
some tens of inverse femtobarns are enough for discovery at the 14
TeV LHC. These studies can be easily adapted to a search for the charged
scalars of our version of the 3-3-1 model and an extended exclusion
region compared to the MSSM case is expected. We will see in the next
section that the associated production of a charged Higgs and a $W$
boson is a more interesting channel to search for the $H_{1}^{\pm}$
states.

\subsection{Associated production of a charged Higgs and a W boson}

In this section we study the associated production of a charged Higgs
and a $W^{\pm}$ boson. The process occurs through the production
of a $Z^{\prime}$ via $q\overline{q}$ annihilation, the $btH_{1}^{\pm}$
Yukawa interaction via $b\overline{b}$ channel, and s-channel neutral
Higgs bosons $h_{1,2}$ diagrams. As a consequence of the custodial
symmetry imposed on the scalar sector of the model and the small difference
between the $u$ and $v$ vacuum expectation values, the $h_{2}$,
$A_{0}$, and $H_{1}^{\pm}$ Higgs bosons are almost degenerated in
mass as we can see comparing the Figs. \ref{fig:zl500}, \ref{fig:zl800},
\ref{fig:zl1500}, and \ref{fig:charged_masses}. As we said above,
$h_{1}$ is equivalent to the SM Higgs boson and is the lightest
state for practically all the parameter space. Therefore, the contribution
from $b\overline{b}\rightarrow h_{1,2}\rightarrow W^{\pm}H_{1}^{\mp}$
is negligible once the neutral Higgs bosons cannot be on their mass shell.

\begin{figure}[H]
 \centering\includegraphics[scale=0.7]{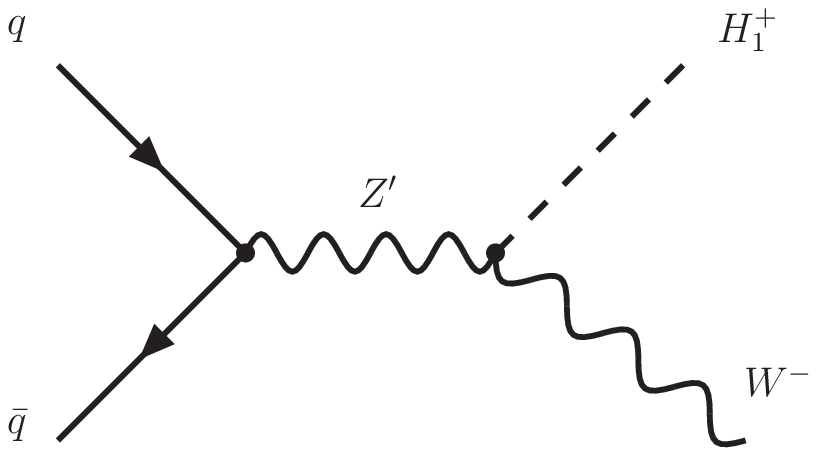}\qquad{}\includegraphics[scale=0.7]{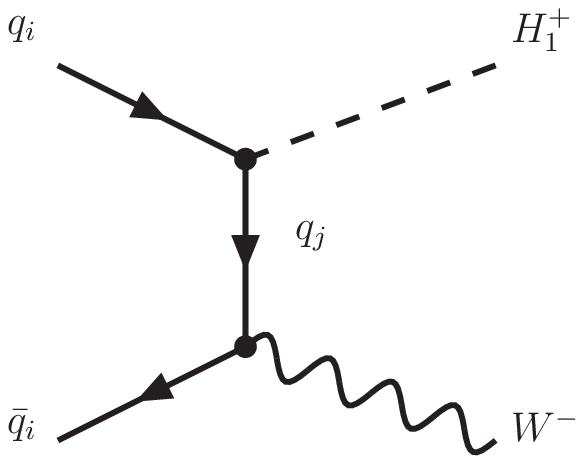}
\caption{Feynman graphs of the contributing processes to $q\overline{q}\rightarrow W^{-}H^{+}$.
At the left panel we show the $Z^{\prime}$s-channel contribution
and at the right panel the heavy quark t-channel diagram from the
Yukawa interaction.{\large \label{fig:feyngraph}}}

\end{figure}

We show in the Fig. \ref{fig:feyngraph} the effective contributing
channels to \textbf{$W^{\pm}H_{1}^{\mp}$} production. The charged
Higgs production through an s-channel $Z'$ is a novel feature predicted
by the model that can lead to very distinctive topologies as we will
see in the next section.

\begin{figure}
\centering\includegraphics[scale=0.6]{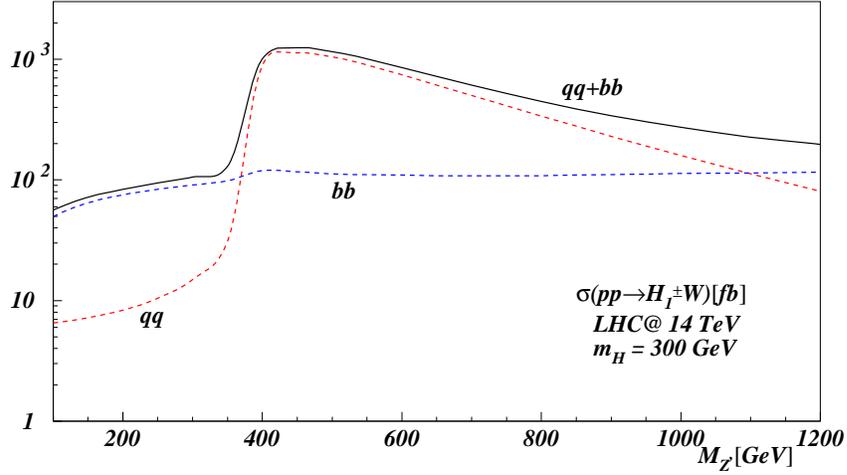}

\caption{Total cross section (in fb) for the associated $H_{1}^{\pm}$ and
$W^{\mp}$ production at the LHC for $\sqrt{s}=14$ TeV as a function
of the $Z^{\prime}$ mass keeping the $H_{1}^{\pm}$ mass fixed at
300 GeV. The dashed lines represent the Yukawa induced process and
the light quark annihilation channel. The solid line is the sum of
the two contributing channels.\label{fig:HW-lhc2}}

\end{figure}

The total cross section as a function of the $Z^{\prime}$ mass, for
a 300 GeV charged Higgs, at the LHC is shown in the Fig. \ref{fig:HW-lhc2}
for $\sqrt{S}=14$ TeV. The bump observed at $m_{Z'}\simeq m_{H_{1}^{\pm}}+m_{W}$ GeV
corresponds to the transition to on shell $Z'$ production via
the $q\overline{q}$ annihilation which is the dominant channel in
that regime.

The t-channel Yukawa production is dominant when either the $Z'$
is off shell or is very heavy as we can see at Fig. \ref{fig:HW-lhc2}.
Near the threshold for on shell production the contribution from the
s-channel $b\overline{b}\rightarrow Z^{\prime}\rightarrow H_{1}^{\pm}W^{\mp}$
increases to about 30\% of the total $b\overline{b}$ channel, but
away from the resonance this contribution is very small.

\begin{figure}
\centering\includegraphics[scale=0.6]{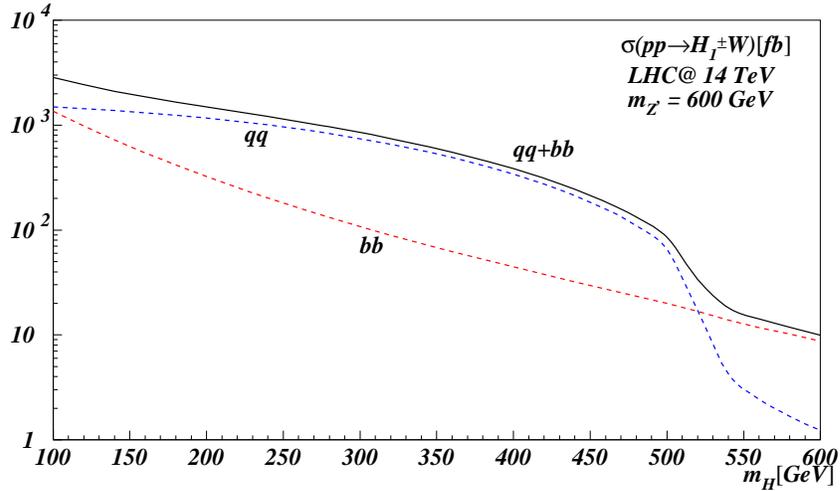}

\caption{Total cross section (in fb) for the associated $H_{1}^{\pm}$ and
$W^{\mp}$ production at the LHC for $\sqrt{s}=14$ TeV as a function
of the charged Higgs mass keeping the $Z^{\prime}$ mass fixed at
600 GeV. The dashed lower line represents the Yukawa induced process,
and the dashed upper line the light quark annihilation channel. The
solid line is the sum of the two contributing channels. \label{fig:HW-ZL600}}

\end{figure}

The sensitivity to the $Z^{\prime}$ mass in the t-channel Yukawa
production enters through the factorization scale that was chosen
as $\mu_{F}=\nicefrac{m_{Z'}}{4}$ for the initial state bottom quarks.
It has been shown \cite{alves-plehn} that choosing a smaller scale
is the appropriated choice in the bottom parton picture of processes
initiated by bottom quarks for the sake of the perturbative stabilization.
This claim is confirmed in this case observing that the $b\overline{b}$
channel varies by a factor of 3 in magnitude in the whole range of
$Z^{\prime}$ masses considered in the Fig. \ref{fig:HW-ZL600}
while the factorization scale varies by a factor of 12. The combined
total cross section including positively and negatively charged states
and $q\overline{q}$ and $b\overline{b}$ channels can be as high
as 1.2 pb for this charged Higgs mass.

QCD corrections are expected to be as small as the Drell-Yan processes
cases and are not included in the analysis. The Yukawa t-channel contributions
may have more substantial QCD corrections \cite{alves-plehn}, but
it is not the dominant contribution as we discussed. Nevertheless
taking into account additional hard radiation may be important for
a proper evaluation of signals and backgrounds in the tail of some
kinematical distributions. We will return to this discussion in the
next section.

The Fig. \ref{fig:HW-ZL600} displays the $q\overline{q}$ and the
$b\overline{b}$ initiated processes separately, and the total $q\overline{q}+b\overline{b}$
production rate as a function of the charged Higgs mass. The initial
state bottom process is important for very light or very heavy charged
Higgs masses. The contribution for a 100 GeV $H_{1}^{\pm}$ reaches
1 pb and decreases by 2 orders of magnitude when $m_{H_{1}^{\pm}}=600$
GeV. This behavior is due mainly to the bottom distribution function
in the proton sea. The $q\overline{q}$ contribution, by its turn,
decreases sharply near the $m_{H_{1}^{\pm}}+m_{W}$ threshold as the
$Z^{\prime}$ gauge boson gets off shell, becoming smaller than the
bottom initiated process.

\begin{figure}
\centering\includegraphics[scale=0.6]{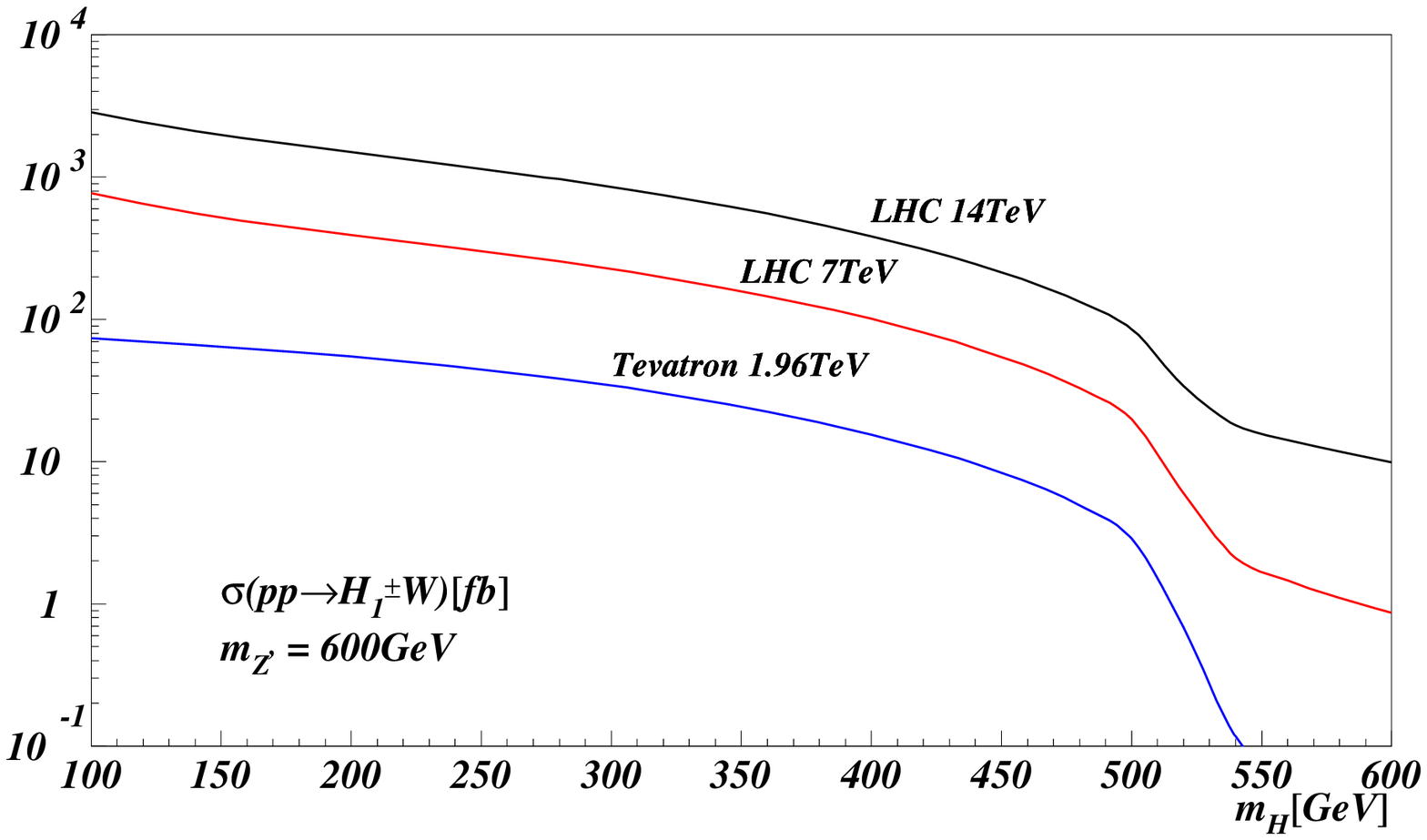}\caption{Total cross sections (in fb) for $W^{\pm}H_{1}^{\mp}$ production
including all the contributing channels. The upper curve shows the
LHC running at 14 TeV, the middle curve the 7 TeV LHC, and the lower
curve the Fermilab Tevatron. The $Z^{\prime}$ mass is held fixed
at 600 GeV. \label{fig:HW-lhc-teva}}

\end{figure}

We show in the Fig. \ref{fig:HW-lhc-teva} the production cross
section as a function of the charged Higgs mass including all s and
t-channels at the 7 and 14 TeV LHC and at the 1.96 TeV Tevatron for
a 600 GeV $Z^{\prime}$ gauge boson. We can see again the threshold
for $Z^{\prime}$ production near $m_{Z'}\simeq m_{H_{1}^{\pm}}+m_{W}$
GeV and the role played by the $b\overline{b}$ channel whose contribution
is important for low charged Higgs masses and in the off shell $Z^{\prime}$
regime and as the center-of-mass energy rises. We clearly see that
Tevatron can only produce a charged Higgs in association with a $W$
boson through an s-channel $Z^{\prime}$. Despite the rates at the
7 TeV LHC can reach hundreds of femtobarns we checked that the reduced
integrated luminosity designed for this run precludes a statistically
significant observation of a charged Higgs boson in the channel under
study. We will return to this discussion later.

This large contribution of the $b\overline{b}$ channel to the total
rates is a feature shared by several nonminimal Higgs sectors extensions
of the SM possessing charged scalars, like the MSSM, for example. Recalling
the interaction Lagrangian for $tbH_{1}^{\pm}$ vertex given by Eq.
(\ref{eq:yuk-h1tb}) we see that the factor $\frac{m_{t}}{v_{w}}\frac{v}{u}$
has a magnitude of order 1 for $m_{t}=174.3$ GeV and the vacuum expectation
values as chosen in the Sec. II. A similar Lagrangian describes
the charged Higgs couplings to $\tau$ leptons and $\tau$ neutrinos.

\subsection{Associated production of a charged Higgs and a top quark}

The initial state bottom induced process $bg\rightarrow tH_{1}^{\pm}\rightarrow t\tau^{\pm}\nu_{\tau}$
is, according to ATLAS~\cite{atlasTDR} and CMS~\cite{cmsTDR}, the
most promising channel for charged Higgs bosons in two-Higgs doublets
models, in particular, for the large $tan\beta$ regime. In part this
is due the large production rates expected for the process at the
LHC including the $tan\beta$ enhancement factor.

\begin{figure}
\includegraphics[scale=0.6]{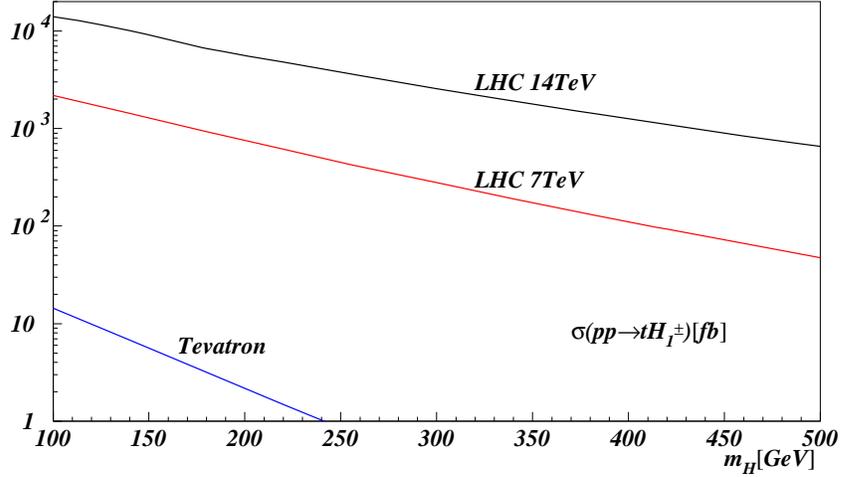}

\centering

\caption{Associated production cross section (in fb) of a top quark and a charged
Higgs boson at hadron colliders as a functions of the scalar mass.
The production of both positive and negative charge states were taken
into account.\label{fig:topHiggs}}

\end{figure}

Similarly the cross section for $bg\rightarrow tH_{1}^{\pm}$ predicted
by the 3-3-1-RHN model benefits the large Yukawa coupling which compensates
the small $b$ quark flux from the proton sea resulting in large rates
as we see in Fig. \ref{fig:topHiggs}. At the LHC running at 14
TeV the total cross section is above the picobarn level for the entire
mass range considered here. This is much larger than the expected
for the MSSM case with $tan\beta=30$ and including NLO QCD corrections
as shown in \cite{topHiggs} except for small Higgs masses or very
large $tan\beta$.

At the LHC running at 7 TeV the cross section is in the hundreds of
femtobarns region for $m_{H_{1}^{\pm}}\lesssim400$GeV while the rates
at the Tevatron are too small as can be seen in the Fig. \ref{fig:topHiggs}
even for small masses.

\subsection{Charged scalars produced in top quark decays}

\begin{figure}
\centering\includegraphics[scale=0.6]{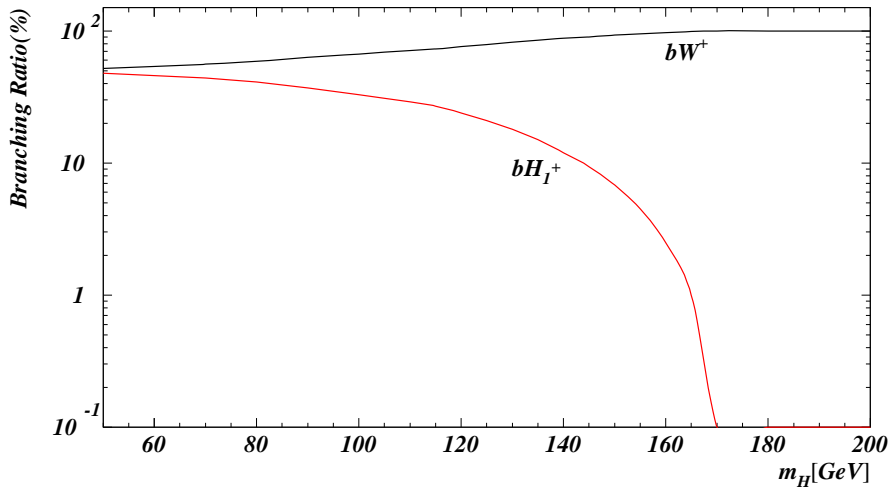}

\caption{The branching ratios of the top quark decays into the
\textbf{$bW^{\pm}$}
and $bH_{1}^{\pm}$ channels as a function of the charged scalar mass.}
\label{fig:brtHCb}
\end{figure}

Another interesting production channel is the top quark pair production
with one top decaying into a bottom quark and a charged Higgs. Of
course, the channel is promising only in the small Higgs mass portion
of the parameter space. We show in the Fig.~\ref{fig:brtHCb} the
branching ratio of
top quarks including the $bH_{1}^{\pm}$channel. The bottom-Higgs
decay can reach a 50\% branching fraction for a 50 GeV Higgs but decreases
very rapidly as the mass increases. However, the branching fraction
is larger than 10\% if $m_{H}\lesssim140$ GeV.

The production cross section for the process $pp\rightarrow t\overline{t}\rightarrow t\overline{b}H_{1}^{+}$
can be easily computed using the branching fraction information. For
example, multiplying the 580 pb $t\overline{t}$ cross section at
LO in the LHC 14 TeV by the branching fraction of 33\% for a 100 GeV
charged Higgs we find a very large rate of 191 pb. This is not a realistic
estimate though because one could choose to tag the bottom quark to
suppress backgrounds and in this case it is necessary to impose acceptance
cuts on the bottom jet. For example, imposing the following acceptance
cuts on the bottom jet

\begin{eqnarray}
p_{T}>20\: GeV\:,\: & \mid\eta_{b}\mid<2.5\label{eq:ptbcut}\end{eqnarray}
 we have a 160 pb exclusive cross section.

Once again, the 3-3-1-RHN model predicts a larger cross section compared
to the MSSM case where, for instance, the branching ratio of a 140
GeV charged Higgs is below the 10\% level for $1\lesssim tan\beta\lesssim30$
and reaches 30\% for $tan\beta\simeq60$ \cite{cmsTDR}. For smaller
masses the region around $tan\beta=7$ still has very small branching
ratios, a well know characteristics of the 2HDM, in special of the
MSSM.

We have just found another distinctive feature of the Higgs sector
of this 3-3-1-RHN model compared to SUSY and doublets models in general:
a possibly sizable number of bottom jets plus tau leptons events associated
to charged Higgs bosons production.

\subsection{Associated production with new neutral gauge bosons}

The production mode $pp\rightarrow Z'H_{1}^{\pm}$ has a negligible
cross section even at the 14 TeV LHC. It is interesting to note that
a $Z'$ decaying to jets and a $H_{1}^{\pm}$decaying taus or to $W^{\pm}h_{1}$
may lead to a bump in the dijet mass as reported by the CDF Collaboration
\cite{cdf-dijets} although with a much lower cross section. Assuming
a 150 GeV $Z'$ and a 100 GeV charged Higgs we found a tiny cross
section of 8 fb only. However there are versions of our model with
a leptophobic $Z^{\prime}$ where this cross section can be much greater
\cite{AAAE2011}.

\section{Phenomenological Analysis for the $H_{1}^{\pm}W^{\mp}$channel}

We demonstrate in this section the potential of the LHC, operating
at 14 TeV center-of-mass energy, to discover the lightest charged
Higgs boson predicted by the model under consideration in the associated
$H_{1}^{\pm}W^{\mp}$ production mode.

Thousands of events are expected based on the production cross sections
computed in the previous section in the low luminosity stages of the
14 TeV LHC accumulating 1 to 10 $fb^{-1}$of data. On the other hand
1000 events at maximum are expected at the 7 TeV LHC after
a 5 $fb^{-1}$run and at the Tevatron after the whole 10 $fb^{-1}$run.
As we will discuss these numbers of events are not enough to claim
a statistically significant discovery after background suppression
based entirely on a cut method. However, an analysis based on the
likelihood ratio statistics may well change this conclusion for the
amount of data to be accumulated at those experiments relying on the
most distinctive kinematical distributions.

\begin{figure}
\includegraphics[scale=0.6]{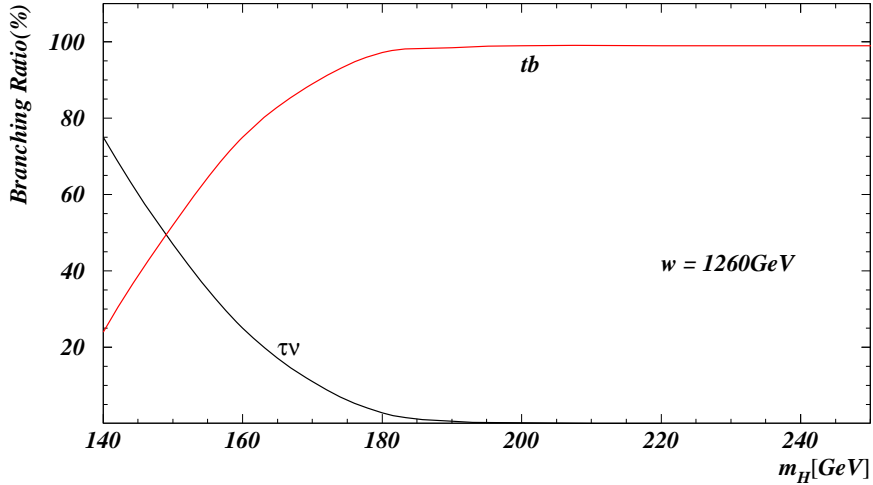}\caption{The branching ratios of the charged Higgs boson as a function of its
mass.\label{fig:Hcbran}}

\end{figure}

\begin{figure}
\includegraphics[scale=0.6]{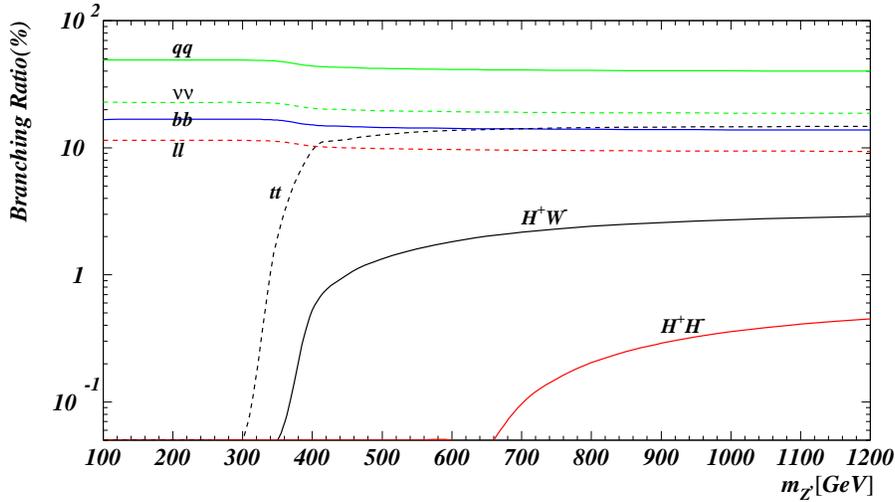}

\caption{The branching ratios of the new neutral gauge boson $Z^{\prime}$
as a function of its mass.\label{fig:Zlbran}}

\end{figure}

The charged Higgs boson decays predominantly to top-bottom pairs for
$m_{H_{1}^{\pm}}>m_{t}+m_{b}$ and to tau leptons, $\tau^{\pm}\nu_{\tau}$,
for smaller masses as we show in the Fig. \ref{fig:Hcbran}. This
is a direct consequence of the Yukawa enhancement factor discussed
in the previous section. The $Z^{\prime}$ gauge boson in its turn
decays most part of the time into light quarks, about 40\% to 50\%,
followed by neutrinos, bottom quarks, and charged leptons (including
$\tau$ leptons). Among the heavy states, the $t\overline{t}$ channel
is the most favored one with a branching ratio close to 15\%. The
associated $H_{1}^{\pm}W^{\mp}$channel has a small branching ratio
around 2\% for Higgs masses from 500 to 1200 GeV as can be seen in
the Fig.~\ref{fig:Zlbran} while the charged Higgs pair decay channel
$Z^{\prime}\rightarrow H_{1}^{\pm}H_{1}^{\mp}$ is very rare. Decays
into heavy new fermions, gauge bosons, and neutral Higgses are negligible
for the range of parameters considered in this work.

Due the large branching ratio into a bottom and a charged Higgs and
the large $t\overline{t}$ production cross section at hadron colliders,
the process

\begin{equation}
pp\rightarrow t\overline{t}\rightarrow b\overline{b}H_{1}^{\pm}H_{1}^{\mp}\rightarrow b\overline{b}\tau^{+}\tau^{-}\nu_{\tau}\overline{\nu}_{\tau}\label{eq:pptt-bbHH}
\end{equation}

 is expected to yield thousands of events at the LHC and the Tevatron
for Higgs masses below the top mass. However, it is not possible to
reconstruct the Higgses because of the escaping neutrinos. Moreover,
the tau lepton decay will dilute considerably the production rates.
Tagging the bottoms and taus could be a way out but the efficiency
for multiple tau lepton and bottom tagging is rather small. Another
possibility would be to identify a top decaying into $bW$ state and
then $W\rightarrow jets$. In this case the reconstruction is possible
at the cost of a increased QCD backgrounds. Anyway, this is a classical
analysis already performed in various previous works on the 2HDM and
MSSM case and the known results can be applied directly to our work
by an appropriate rescaling. We postpone this study to a future work.

The associated production of a top quark and the charged Higgs has
the second largest cross section at the LHC as we have computed in
the previous sections. Taking into account the preferred decay mode
of a heavy charged Higgs, the final state configuration would be

\begin{equation}
pp\rightarrow tH_{1}^{\pm}\rightarrow ttb\rightarrow bbb+W^{+}W^{-}\label{eq:pptH-ttb}\end{equation}
 which could lead to a many jets plus lepton configuration or a cleaner
dilepton configuration. The cleaner configuration involves two missing
neutrinos which precludes the reconstruction of the charged Higgs
resonance. The many jets configuration is expected to have a higher
level of QCD fakes, but the reconstruction is possible. A study of
this channel is currently underway \cite{AAAE2011}.

Despite the smaller cross section, the associated production of a
$W$ boson and the charged Higgs present a very distinctive signal

\begin{equation}
pp\rightarrow W^{\pm}H_{1}^{\mp}\rightarrow W^{\pm}tb\rightarrow b\overline{b}+W^{+}W^{-}\label{eq:ppWH-Wtb}\end{equation}
 where the $W$ bosons may decay into leptons or jets. The dilepton
configuration is the cleaner way to search for the charged Higgs,
but again we loose the Higgs resonance. The mixed jets plus lepton
state, on the other hand, permits the reconstruction up to a twofold
ambiguity in the neutrino momenta. Moreover, the chain of resonances
endows the signal very singular features that allow us to separate
it from the SM backgrounds. The new ingredient here is the presence
of the $SU(3)_{L}$ neutral gauge boson in the s-channel that decays
to a $W^{\pm}H_{1}^{\mp}$ pair. The presence of this $Z^{\prime}$
not only increases the production rates but makes the leptons and jets
harder than the expected from processes induced solely by t-channel
Yukawa interactions as in the usual two-Higgs doublets models.

The signal to be studied in this work is the following

\begin{equation}
pp\rightarrow W^{\pm}H_{1}^{\mp}\rightarrow b\overline{b}W^{+}W^{-}\rightarrow b\overline{b}+jj+\ell\nu_{\ell}\label{eq:ppWH-bbWW}\end{equation}
 where the light jets and leptons come from a $W$ boson decay and
$\ell$ denotes an electron or muon. In order to avoid the huge QCD
backgrounds we propose a double $b$-tagging assuming a 60\% $b$-tag
efficiency and a $5\times10^{-3}$ rejection factor against mistagged
light quark and gluon jets \cite{btag}. We assume a 90\% efficiency
for lepton identification and include a Gaussian smearing of the energy
of jets and the lepton but not for their momentum direction.

It is important to describe more carefully the chain of resonances
in order to understand our search strategy. After production, the
charged Higgs boson decays to a top and bottom pair

\begin{equation}
W^{\pm}[H_{1}^{\mp}\rightarrow tb]_{A}\label{eq:}\end{equation}
 the top quark in its turn will decay into another $W$ boson and
$b$ quark, and we will have

\begin{equation}
W^{\pm}[H_{1}^{\mp}\rightarrow(t\rightarrow W^{\mp}b)_{B}b]_{A}\label{eq:-1}\end{equation}
 The $W$ bosons will decay at last producing the visible particles
to the detector and a neutrino, so we get at the end of the whole
decay chain

\begin{equation}
(W^{\pm}\rightarrow\ell^{\pm}\nu_{\ell})_{D}[H_{1}^{\mp}\rightarrow(t\rightarrow(W^{\pm}\rightarrow jj)_{C}b)_{B}b]_{A}\label{eq:-2}\end{equation}

There are four resonant states that we labeled as $A,B,C$, and $D$.
There is, of course, the primary $Z^{\prime}$ resonance; however
we will not try to reconstruct the new gauge boson in our phenomenological
analyses just because our main task is to unravel the presence of
the charged Higgs boson. A separate work dedicated to the goal of
identifying the new gauge bosons predicted by the 3-3-1RHN model is
something important by itself and will not be addressed here.

The second step to explain the search strategy is listing the Standard
Model backgrounds involving QCD and electroweak (EW) interactions
that could mimic our signal:
\begin{enumerate}
\item QCD+EW+$Z^{\prime}$ $t\overline{t}$ production is the most important
irreducible background
\item QCD+EW $b\overline{b}+jj+W$ including SM $Z$ bosons, photons and
SM $W$ bosons decaying to bottoms and jets
\item Single top quark production $tbW$
\item QCD+EW $jjjj+W$ with mistagged light quark/gluon jets.
\end{enumerate}
Except for the top quark pair production, where the
factorization/renormalization
scale was chosen to be $\mu_{F}=\mu_{R}=m_{t}$, all the other backgrounds
were computed with an event-by-event factorization/renormalization
scale defined by the square root of the combined transverse momentum
of the identified jets, $\mu_{F}=\mu_{R}=\sqrt{\sum_{jets}\, p_{T}^{2}}$.

\begin{figure}
\begin{raggedright} \includegraphics[scale=0.59]{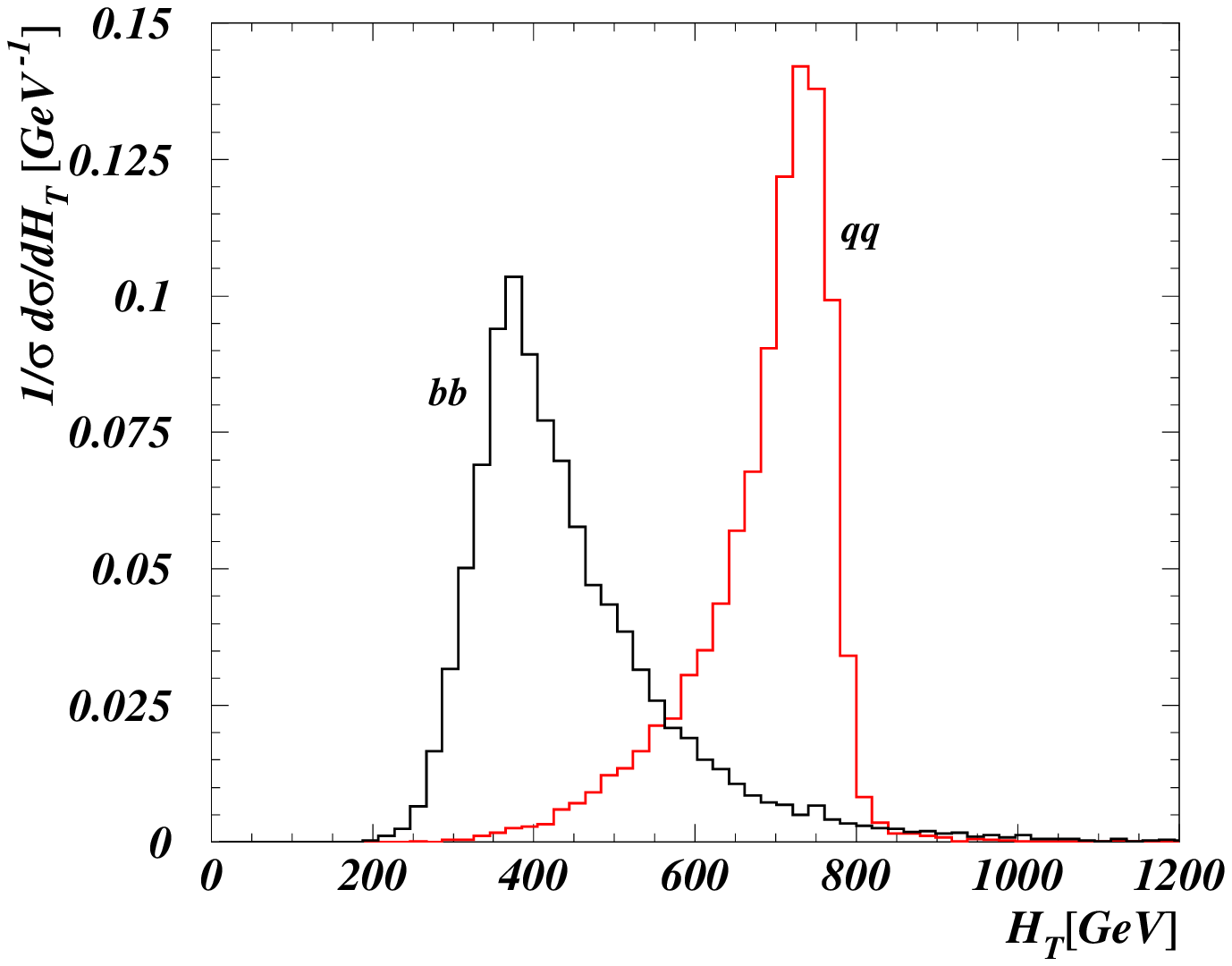}\includegraphics[scale=0.59]{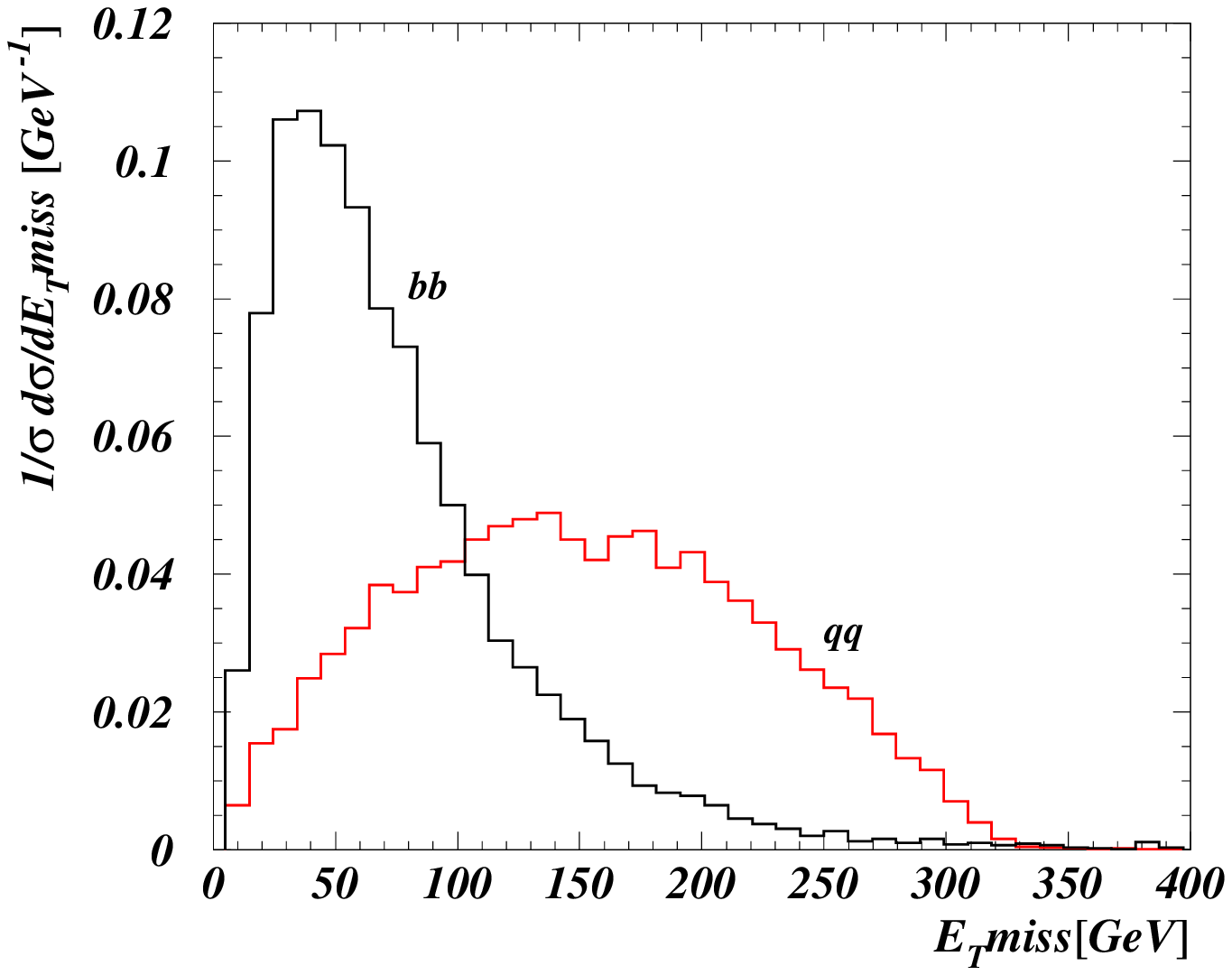}

\end{raggedright}

\begin{raggedright} \includegraphics[scale=0.59]{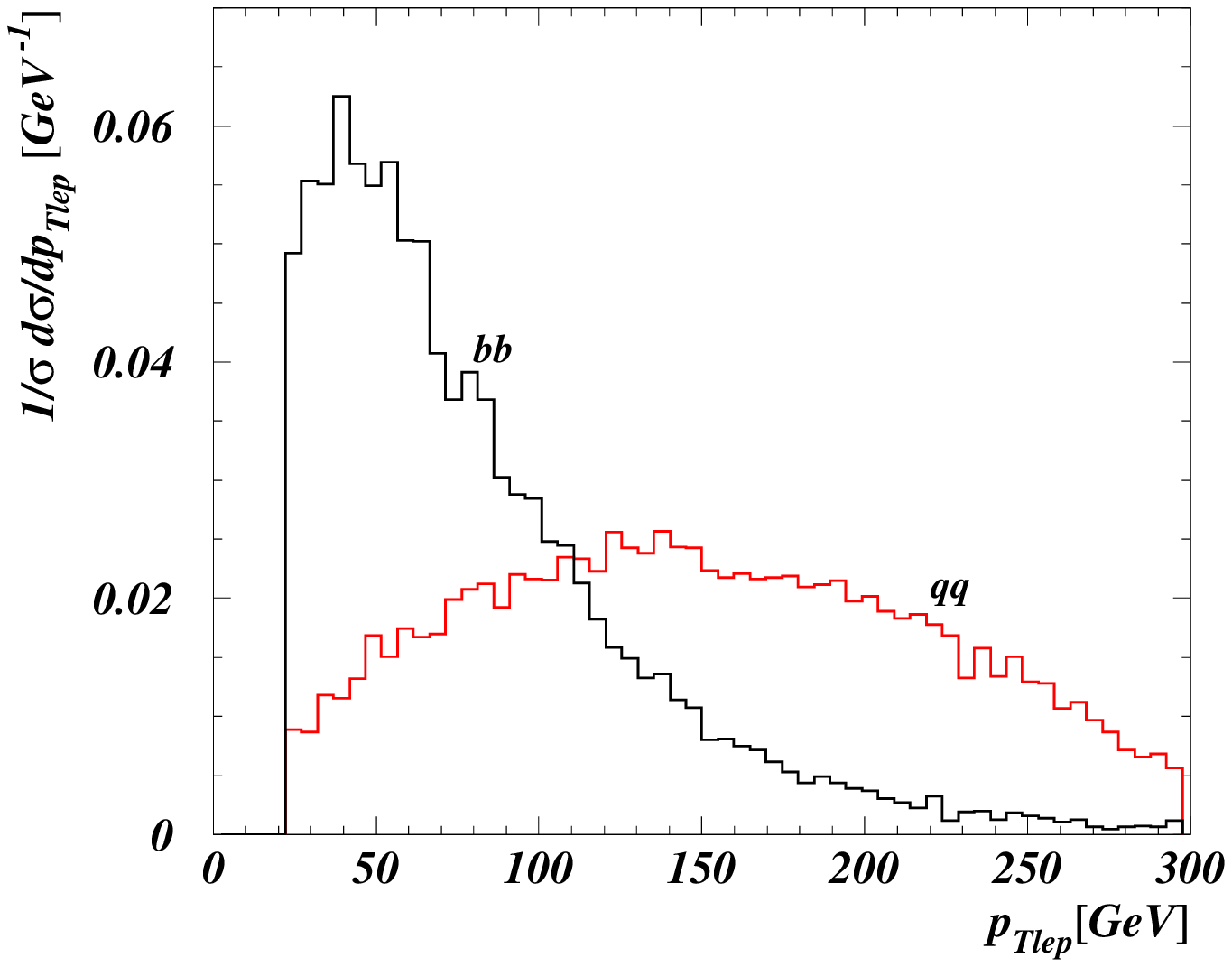}

\end{raggedright}

\caption{The left upper panel, the right upper panel, and the lower panel display
the $H_{T}$ of the defined jets, the missing $E_{T}$ due the escaping
neutrinos, and the transverse momentum $p_{T}$ of the charge lepton,
respectively, for the signal particles produced in the $q\overline{q}\rightarrow Z^{\prime}\rightarrow W^{\pm}H_{1}^{\mp}$
and $b\overline{b}\rightarrow W^{\pm}H_{1}^{\mp}$ with a t-channel
top. The distributions are normalized by the total cross section.
The mass of the $Z^{\prime}$and the charged Higgs are 800 GeV and
300 GeV, respectively. \label{fig:bb_qq}}

\end{figure}

For all backgrounds and the signal we imposed the following acceptance
cuts :

\begin{eqnarray}
p_{Tj,b}>30\: GeV\;,\; & p_{T\ell}>100\; GeV\label{eq:ptjbptl}\end{eqnarray}

\begin{eqnarray}
\mid\eta_{j,b,\ell}\mid<2.5\;,\; & \Delta R_{ik}>0.4\;,\; i,k=j,b,\ell\label{eq:etaR}\end{eqnarray}
 the high $p_{T}$ cut on the leptons is a good discriminant between
signal and backgrounds and an excellent experimental trigger too.
On the other hand it favors the signal events resulting from the s-channel
$q\overline{q}\rightarrow Z^{\prime}$ boson over the t-channel Yukawa
contributions initiated by bottom quarks as can be seen in the Fig.
\ref{fig:bb_qq}. For this plot $m_{Z'}=800$ GeV and $m_{H^{\pm}}=300$
GeV.

After acceptance cuts the signal cross section is still deeply buried
beneath the backgrounds as can be read from the Table \ref{tab:Cuts}
that shows the effect of cuts on the $m_{Z'}=800$ GeV and $m_{H^{\pm}}=300$
GeV signal and backgrounds. To further suppress the backgrounds we
impose a second set of cuts exploring the fact that our signal events
can be much harder than the backgrounds depending on the $Z^{\prime}$
mass

\begin{eqnarray*}
H_{T}>500\: GeV\;,\; & E_{Tmiss}>60\: GeV
\end{eqnarray*}
 where $H_{T}=\sum\, p_{T}$ of all hadrons. In fact the $Z'$ events
are also much harder than the t-channel Yukawa contributions as we
observe in the Fig. \ref{fig:bb_qq}. At this point the t-channel
bottom initiated process contributes only 3\% of the total rates.
Based on this analysis we can understand why the MSSM analogue process
is not a good search channel for charged Higgses. In the MSSM and
2HDM in general, the $W^{\pm}H_{1}^{\mp}$ final state can be produced
via t-channel Yukawa interactions to heavy quarks and s-channel neutral
Higgses contributions both initiated by bottom quarks from gluon splittings.
As a result only charged Higgses decaying to tau leptons in the large
$tan\beta$ regime can be detected at a $5\sigma$ significance level
\cite{WHPRO5}.

\begin{figure}
\centering

\includegraphics[scale=0.6]{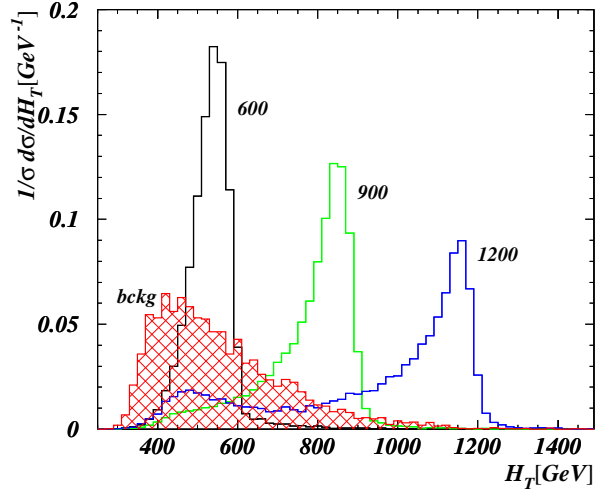}

\caption{The normalized $H_{T}$ distributions for the defined jets from signal
and background events. The signal distributions are shown for a $Z^{\prime}$
gauge boson mass of 600, 800, and 1200 GeV, respectively.\label{fig:HT} }

\end{figure}

The $H_{T}$ distribution for signal and backgrounds from Fig. \ref{fig:HT}
confirms our expectation that heavier $Z^{\prime}$ bosons are much
easier to separate from backgrounds. In the subsequent analyses we
do not try to optimize the cuts to take advantage of these features
instead we would rather keep the analyses as independent of the particular
parameter space point as possible. It is important to point out though
that several improvements can be embodied in a more complete analysis.

Even after imposing such hard set of cuts on the candidate events,
the backgrounds are big enough to preclude any significant conclusion
as can be read from the second row of Table \ref{tab:Cuts}.

The combined QCD+EW+$Z^{\prime}$ $t\overline{t}$ production rate
at the 14 TeV LHC is around 580 pb, almost 3 orders of magnitude
larger than our signal. The need to suppress those backgrounds led
us to choose searching for the resonance $A$, associated to the charged
Higgs decay, into a purely hadronic channel. We did not take into
account the semileptonic top quarks for our signal. Despite we loose
half of our total number of signal events, this give us a tool to
explore an important advantage: the signal contains only one final
state top quark.

The possibility to reconstruct the hadronic and the semileptonic
tops from the $t\overline{t}$ background gives us the opportunity
to tag the semileptonic tops and reject them. This can be easily
done demanding a cut on the bottom-lepton invariant mass. First of
all we look for the bottom quark from the hadronic top and jets from
a $W$ boson, let us call it the first bottom quark, then we impose
the following invariant mass constraints

\begin{eqnarray}
\mid m_{jj}-m_{W}\mid<20\: GeV\;,\; & \mid m_{jjb}-m_{t}\mid<20\: GeV\label{eq:-3}\end{eqnarray}

The first one selects jets from $W$ boson decays and helps to clean
the QCD backgrounds with at least two final state jets. The second
one rejects all processes not related to top quark decays. Until this
point we have made use of two resonance structures from Eq. \ref{eq:-2},
namely, $B$ and $C$.

\begin{figure}
\centering\includegraphics[scale=0.6]{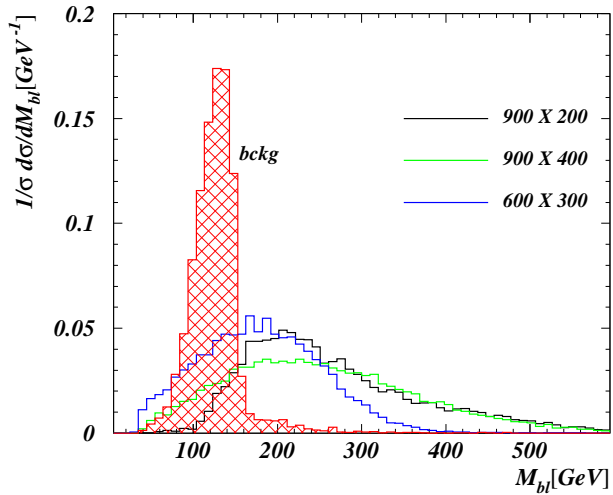}

\caption{The second bottom (as defined in the text) and the charged lepton
invariant mass distribution of the signal and the combined backgrounds.
At this point the backgrounds are still dominated by the $t\overline{t}$
contribution. The distributions are normalized by the total cross
section. The mass $Z^{\prime}$ and the charged Higgs are shown in
the figure. \label{fig:Mbl}}

\end{figure}

The second signal bottom quark that comes from the charged Higgs decay
is not correlated to the lepton from $W$, whereas the second $b$
quark from the $t\overline{t}$ background is the yield of the top
quark that decays semileptonicaly. Thus we expect that the $t\overline{t}$
background events show and end point structure in the invariant mass
of second bottom quark and the charged lepton $m_{bl}<\sqrt{m_{t}^{2}-m_{W}^{2}}\cong160\: GeV$
as can be seen in the Fig. \ref{fig:Mbl}.

Imposing the cut

\begin{equation}
m_{bl}>180\: GeV
\end{equation}

 eliminates almost all the $t\overline{t}$ events while diluting
the signal by a factor of $0.69$ only for our benchmark example.
We clearly see that heavier $Z^{\prime}$ bosons and lighter charged
Higgses present the harder spectrum and are favored by our analysis.
We show in the fourth row of the Table \ref{tab:Cuts} the impact
on the dominant $t\overline{t}$ background and the signal. We can
also observe the large impact of this cut on the EW $WWZ+WZZ$ backgrounds.

The requirement of on shell production of $W$ bosons and hadronic
top quarks plus the tagging of the semileptonic tops is very effective
against SM backgrounds from top pairs and electroweak gauge boson
production. However there is another source of background events which
is not so severely affected by those cuts. The single top process
has a topology similar to our signal: one single top decaying to hadrons,
and a second bottom and a charged lepton not correlated to a top quark
decay. This similarity turns the single top background the dominant
one after the semileptonic top veto. The $79.1$ fb background that
we read from the second column of Table \ref{tab:Cuts} is the due
almost entirely the contribution of the single top process.

\begin{figure}
\centering\includegraphics[scale=0.6]{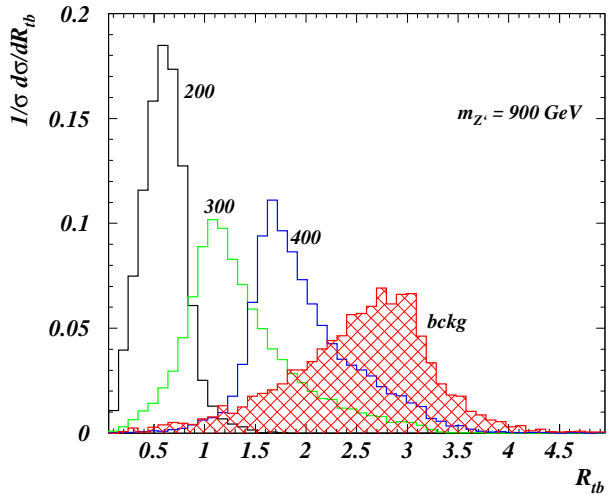}

\caption{The normalized distribution on the distance between the hadronically
tagged top quark and the second bottom jet for three different Higgs
masses and a $Z^{\prime}$ mass of 800 GeV, and the combined backgrounds.
The single top channel is the dominant contribution after the semileptonic
top veto.\label{fig:Drtop}}

\end{figure}

If the $Z^{\prime}$ boson is much heavier than the charged Higgs
we should expect that a high boost will collimate the hadronic top
and the second bottom quark (originated from the Higgs decay) into
a fat jet. That is what we precisely observe in the Fig. \ref{fig:Drtop}
which displays the distance distribution between the hadronic tagged
top and the second bottom, $\Delta R_{tb}$. As the difference $m_{Z'}-m_{H_{1}^{\pm}}$
increases the Higgs jet gets narrower, whereas the top-bottom pairs
from the single top background are much more separated. To suppress
single tops we impose an additional cut on the $\Delta R_{tb}$ variable

\begin{equation}
\Delta R_{tb}<1.6\end{equation}

This cut efficiently dilutes the remaining backgrounds by a factor
of $0.15$, whereas its impact on the signal is small for our benchmark
point. It should be pointed out though that this cut may impact much
more strongly the signal if the difference between the $Z^{\prime}$
and the charged Higgs masses decreases.
For heavy masses this negative
impact can be compensated by hardening the transverse momentum cuts.

We also checked that there exists many good discriminant distributions
that could be used to disentangle signal from backgrounds in specific
points of the parameters space. We did not try to optimize our analysis
for specific regions; instead we show that requiring only trigger cuts
planned to capture possible new physics signals and exploring tagging
techniques for SM and new particles is just enough to observe the
charged Higgses of the model.

\begin{table}
\begin{tabular}{|c|c|c|c|c|}
\hline
Cuts  & Signal  & $t\overline{t}+$ single top  & $Wb\overline{b}jj$  & $WWZ+WZZ$\tabularnewline
\hline
\hline
Acceptance  & 17.3  & $12.624\times10^{3}$  & 734.4  & 6.50\tabularnewline
\hline
$H_{T}>500\: GeV\,,\, E_{T,miss}>60\: GeV$  & 13.7  & $2.807\times10^{3}$  & 196.2  & 2.15\tabularnewline
\hline
$\mid m_{jj}-m_{W}\mid<20\: GeV\,,\,\mid m_{jjb}-m_{t}\mid<20\: GeV$  & 13.4  & $2.612\times10^{3}$  & 5.91  & 1.42\tabularnewline
\hline
$m_{bl}>180\: GeV$  & 9.26  & 79.1  & 3.92  & 0.12\tabularnewline
\hline
$\Delta R_{tb}<1.6$  & 7.62  & 11.5  & 2.74  & $<10^{-2}$\tabularnewline
\hline
$\mid m_{had}-m_{H^{\pm}}\mid<20\: GeV$  & 7.00  & 2.07  & 0.38  & $<10^{-3}$\tabularnewline
\hline
\end{tabular}

\caption{The effect of the various levels of cuts devised to separate the signal
from a 800 GeV $Z^{\prime}$decaying into a 300 GeV $H_{1}^{\pm}$
and a $W$ boson from the standard model backgrounds. In the third
column is the QCD background, and in the fourth and fifth columns
are the electroweak backgrounds. The $b$ tagging and lepton efficiencies
were not taken into account yet. Gaussian smearing of energies and
momenta (only the magnitude, not directions) are included in all rows.\label{tab:Cuts}}

\end{table}

The $W+4j$ where all jets originate from QCD radiation is huge after
the jets acceptance cuts reaching several nanobarns. Assuming a bottom
miss-tagging factor against light quark and gluons jets of $5\times10^{-3}$,
the size of this background drops to the tens of femtobarns level. We checked
that imposing the additional cuts virtually eliminates this source
of backgrounds. Our simulations were performed at parton level though,
and a more realistic computation including hadronization and showering
and detector efficiencies will be necessary to confirm this claim;
however we believe that the mass shell constraints are tight enough
to clean the $W+4j$ events.

Before looking for the charged Higgs resonance, a $S/B=0.54$ after
applying all cuts can be achieved. A clear resonance in the $jjb\overline{b}$
invariant mass corresponding the production of a charged Higgs boson
and subsequent decay to a top-bottom pair is visible over the total
backgrounds for moderate Higgs masses from 200 to 300 GeV at least
as we show in the Fig. \ref{fig:Inva_mass} below. The three signal
lines in the 300 GeV bin represent events from three different $Z^{\prime}$
masses: 800, 1000, and 1200 GeV. The peak from the 200 GeV Higgs is
the more pronounced but lies in a region richer in background events.
The 400 GeV resonance is the less pronounced mainly because of the
$\Delta R_{tb}$ cut devised to eliminate the single top backgrounds.
Despite the smaller phase space volume to produce heavy states, the
300 GeV resonance is more favored than the 200 GeV line even for very
heavy $Z^{\prime}$ bosons once they yield harder jets and leptons
which are more likely to pass the kinematical cuts.

\begin{figure}
\centering\includegraphics[scale=0.6]{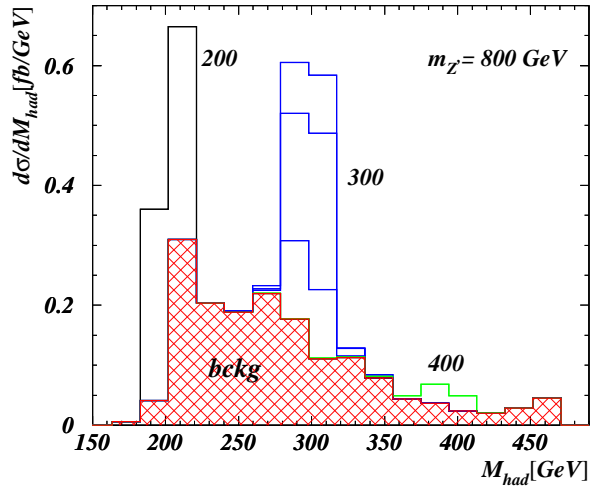}

\caption{The invariant mass of the identified light quark and gluon jets plus
the tagged bottom jets from the charged Higgs decay for the total
backgrounds and the signal plus backgrounds. The empty histograms
show the 200, 300, and 400 GeV Higgs boson resonances for a 800 GeV
$Z^{\prime}$ boson. For the 300 GeV Higgs boson we also show the
distributions for a 1 and 1.2 TeV new neutral gauge boson represented
by the lower lines. \label{fig:Inva_mass}}

\end{figure}

Based on this distribution we compute the required luminosity for
a statistically significant identification of the resonance associated
to a charged Higgs boson production, looking for an excess of events
in a window of 20 GeV around the invariant mass of the identified
jets which we call $M_{had}$. Recalling Eq. \ref{eq:-2}, $M_{had}$
is nothing but the invariant mass associated to the resonant structure
$A$ from the charged Higgs boson decay.

Before the discussion of the LHC potential to search for these charged
scalars we would like to emphasize that a more complete work including
hadronization, showering, and extra radiation is needed to confirm
our claims and estimate the impact of extra jets in the reconstruction
of the Higgs resonance. Moreover, the NLO QCD effects including hard
jet emission for signal and backgrounds would be an important improvement
once the resonances are expected to show up in the tail of jets invariant
mass distribution.

On the other hand some improvements can be devised
in order to separate the signal and backgrounds even more efficiently,
for example, if already exists a hint about the $Z^{\prime}$ mass
scale. In this case, as we have seen, if the neutral gauge boson is
heavy, is possible to impose much harder cuts on jets and leptons.
In special, we checked that bottom transverse momentum is a good discriminant
for very heavy states; however it must be kept in mind that hardening
the bottom cuts could drastically decrease the $b$-tagging efficiency
\cite{btag}. Moreover, it is easy to incorporate the signal semileptonic
top quarks simply vetoing the hadronically decaying ones. In this
case, the reconstructed charged scalars would suffer from a twofold
ambiguity due the two momenta solutions for the neutrino momentum,
but this is not an issue at all.

\begin{figure}
\centering\includegraphics[scale=0.6]{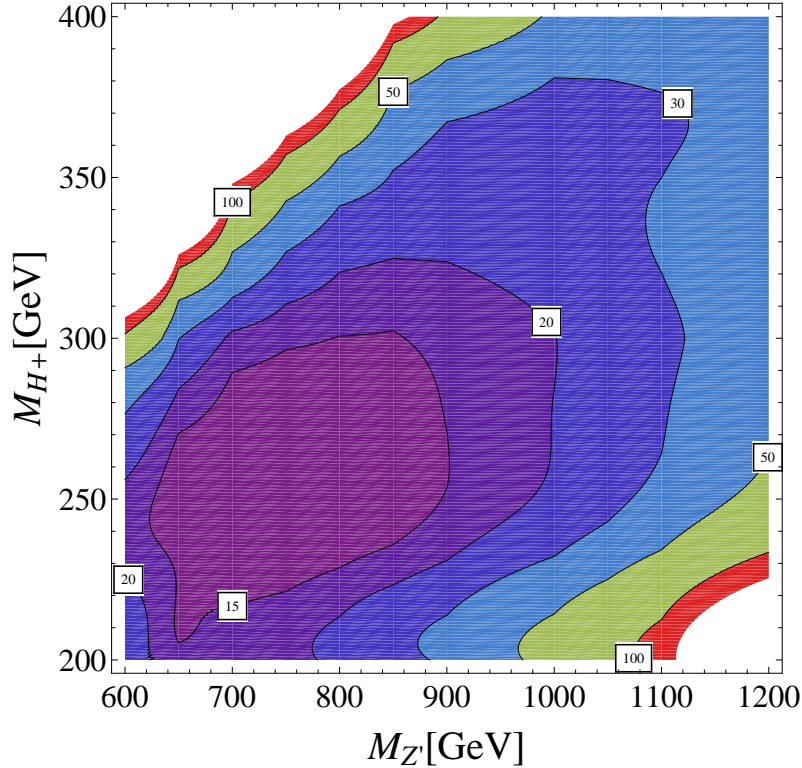}

\caption{The required integrated luminosity for a $5\sigma$ level statistical
significant confirmation of the signal hypothesis over the background
hypothesis in the $m_{Z'}\times m_{H_{1}^{\pm}}$ plane. For this
analyses we assumed $\lambda_{2}=0.31$. \label{fig:scan_MHZl}}

\end{figure}

The Fig. \ref{fig:scan_MHZl} shows the integrated luminosity required
for a $5\sigma$ significance level observation based on the $S/\sqrt{S+B}$
statistics in the $m_{Z'}\times m_{H_{1}^{\pm}}$ plane for $\lambda_{2}=0.31$
fixed based on the $M_{had}$ distribution shown at Fig. \ref{fig:Inva_mass}.
The lower left corner of the masses plane $[600\leq m_{Z'}\leq900]\times[200\leq m_{H^{\pm}}\leq300]$
is the easier place for discovery as a consequence of the enhanced
production cross section. A large portion of this corner demands only
$15\: fb^{-1}$ of data at most for a $5\sigma$ observation. The
upper left corner $[600\leq m_{Z'}\leq900]\times[300\leq m_{H^{\pm}}\leq400]$,
by its turn, is the less favored region for discovery mainly because of
the $\Delta R_{tb}$ cut imposed to eliminate the single top backgrounds
as we discussed earlier. A large portion of this corner cannot be
probed even for $100\: fb^{-1}$ or more.

In the rest of this parameters space we observe the encouraging tendency
to observe the heavier states. This is a direct consequence of the
fact that heavy $Z^{\prime}$ bosons yield hard jets and leptons which
are much more likely to pass cuts compared to the background events
even considering the phase space suppression. It is worth noting again
that the presence of a new gauge boson coupling to $W^{\pm}H_{1}^{\mp}$
pairs, which is responsible for these singular kinematical configurations,
is a distinguishing feature of the model as compared to MSSM and general
2HDM.

\begin{figure}
\centering\includegraphics[scale=0.6]{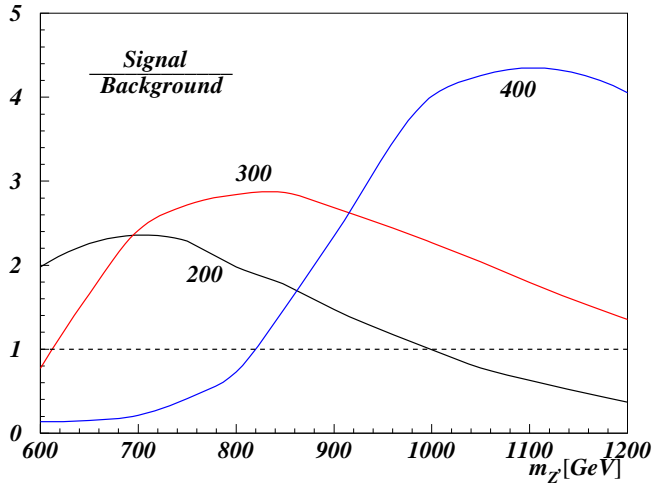}

\caption{The $S/B$ ratio as a function of the $Z^{\prime}$ mass for three
different charged Higgs masses: 200, 300, and 400 GeV.\label{fig:sigback}}

\end{figure}

In the Fig. \ref{fig:sigback} right below we show the $S/B$ ratio
as a function of the $Z^{\prime}$ mass for charged Higgses of 200,
300, and 400 GeV. The intermediate 300 GeV Higgs bosons yield a $S/B>1$
for almost all $Z^{\prime}$ masses considered in this work. The 200
and 400 GeV Higgses show a $S/B>1$ for light and heavy $Z^{\prime}$,
respectively, and this is a consequence of our more or less blind
set of cuts. As we discussed earlier the $\Delta R_{tb}$ cut favors
large $m_{Z'}\times m_{H_{1}^{\pm}}$ regions, but a more dedicated
analysis can be made in order to observe the heavy Higgs and the light
$Z^{\prime}$ portions of the parameters space. Note that the 400
GeV case presents the larger $S/B$ ratios for heavy new gauge boson masses
because of the much harder jets, bottoms, and leptons from their decays
are much more likely to pass the cuts. On the other hand, the reduced
production cross section demands more accumulated data for a significantly
statistical observations as can be seen in Fig. \ref{fig:scan_MHZl}.

As a final remark we point out that $S/B\ge 1$ is a robust prediction in the
sense it is less sensitive to systematic errors.

\section{Conclusions}

If the LHC collaborations could find experimental evidence for an
scalar charged particle this will be an undisputed evidence of new
physics beyond the SM. Moreover, it will shed light on the mechanism
of electroweak symmetry breaking. In this work, we analyzed the scalar
spectrum of the 3-3-1 model with right-handed neutrinos but with a
custodial symmetry that reduces the number of parameters in the scalar
potential. There are two charged and five neutral scalar states in
the particle spectrum. The lightest charged Higgs bosons, $H_{1}^{\pm}$,
can be as light as 100 GeV.

The model presents several distinguishing features associated to the
new gauge bosons and scalars. In special, there is an important interplay
between one of the new neutral gauge bosons, $Z^{\prime}$, and the
charged scalars of the model that enhances the production cross sections
as compared to the MSSM or 2HDM. Yukawa interactions between the charged
Higgses and heavy quarks are strong, which results in large cross sections
for the typical production modes as the associated top-charged Higgs
and charged Higgses from top quark decays. The charged Higgs pair
production is also expected to be larger than its MSSM and 2HDM analogues
as a consequence of the enhanced Yukawa interactions and the contribution
from the new $Z^{\prime}$ gauge boson.

As an example of the role played by this new $SU(3)_{L}$ neutral
gauge boson, the light charged Higgs bosons $H_{1}^{\pm}$ can be
produced in association to a SM $W$ boson through the $Z^{\prime}$
decay. The charged Higgses decay into top and bottom quarks for $m_{H_{1}^{\pm}}>m_{t}+m_{b}$,
which lead to a $tbW\rightarrow b\overline{b}jj\ell\nu_{\ell}$ final
state at hadron colliders as the LHC and the Tevatron. The top quark,
the $W$ boson, and the charged Higgs resonances make the task to
separate the signal from backgrounds relatively simple, and as a consequence
a large portion of the parameters space of the model will be accessible
to the 14 TeV LHC Collaborations with up to $50\: fb^{-1}$, and even
a $15\: fb^{-1}$ integrated luminosity would be just enough to discover
a charged Higgs boson. We found that the same conclusion can not be
claimed for the 7 TeV LHC and the Tevatron because of the reduced amount
of data designed for these experiments. Nevertheless the lighter charged
Higgs, $m_{H_{1}^{\pm}}<m_{t}+m_{b}$, decays predominantly to tau
leptons and if the $Z^{\prime}$ is not too heavy this channel might
be a good search channel for the Tevatron or even a 7 TeV LHC in a
longer run.

We did not try to optimize our analyses for very specific points of
the parameters space, instead we focused on a more or less blind set
of cuts taking into account only acceptance, trigger and tagging techniques.
The robustness of the analyses shows up as signal to background ratio
greater than 1 for a large portion of the parameters space under study
in this work. Detailed studies can be made though in order to take
advantage of the hard jets and leptons expected for even heavier $Z^{\prime}$
and charged Higgses.

\paragraph*{Acknowledgments: }

E. R. B. thanks PNPD-Capes and FAPESP for the financial support. A.G.D.
also thanks FAPESP and CNPq for the financial support.

\appendix

\section{Relevant Interaction Terms\label{sec:ApA}}

$H_{1}^{+}W^{-}h_{1}$ \begin{equation}
\mathcal{L}=-\frac{i\sqrt{2}g}{4}\frac{\left(u-v\right)}{v_{w}}\left[h_{1}^{0}\overleftrightarrow{\partial_{\mu}}H_{1}^{+}\right]W^{\mu-}+H.c\label{eq:H1Wh1}\end{equation}

$H_{1}^{+}W^{-}h_{2}$\begin{equation}
\mathcal{L}=\frac{i\sqrt{2}g}{4}\frac{\left(u+v\right)}{v_{w}}\left[h_{2}^{0}\overleftrightarrow{\partial_{\mu}}H_{1}^{+}\right]W^{\mu-}+H.c\label{eq:H1Wh2}\end{equation}

$H_{1}^{+}H_{1}^{-}A_{\mu}$ \begin{eqnarray}
\mathcal{L} & = & -ie\left[H_{1}^{+}\overleftrightarrow{\partial_{\mu}}H_{1}^{-}\right]A^{\mu}\label{eq:H1H1A}\end{eqnarray}

$H_{1}^{+}H_{1}^{-}Z_{\mu}$ \begin{eqnarray}
\mathcal{L} & = & -\frac{ig}{2\mathrm{cw}}\left(1-2\mathrm{sw^{2}}\right)\left[H_{1}^{+}\overleftrightarrow{\partial_{\mu}}H_{1}^{-}\right]Z^{\mu}\label{eq:H1H1Z}\end{eqnarray}

$H_{1}^{+}H_{1}^{-}Z_{\mu}^{\prime}$ \begin{eqnarray}
\mathcal{L} & = & -\frac{ig}{6v_{w}^{2}\sqrt{t^{2}+3}}\left(2\left(2v^{2}+u^{2}\right)t^{2}-3\left(u^{2}-v^{2}\right)\right)\left[H_{1}^{+}\overleftrightarrow{\partial_{\mu}}H_{1}^{-}\right]Z^{\prime\mu}\label{eq:H1H1Zl}\end{eqnarray}

$H_{1}^{+}W_{\mu}^{-}Z^{\prime\mu}$\begin{eqnarray}
\mathcal{L} & = & \frac{g^{2}uv}{v_{w}}\frac{\mathrm{cw}}{\sqrt{3-4\mathrm{sw^{2}}}}H_{1}^{+}W^{\mu-}Z_{\mu}^{\prime}\label{eq:H1WZl}\end{eqnarray}

$H_{2}^{+}H_{2}^{-}Z_{\mu}$\begin{eqnarray}
\mathcal{L} & = & -\frac{ig}{2\mathrm{cw}}\left[\frac{u^{2}}{\left(u^{2}+w^{2}\right)}-2\mathrm{sw^{2}}\right]\left[H_{2}^{+}\overleftrightarrow{\partial_{\mu}}H_{2}^{-}\right]Z^{\mu}\label{eq:H2H2Z}\end{eqnarray}

$H_{2}^{+}H_{2}^{-}Z_{\mu}^{\prime}$\begin{eqnarray}
\mathcal{L} & = & \frac{ig}{2\mathrm{cw}}\frac{\left(1-2\mathrm{sw^{2}}\right)}{\sqrt{3-4\mathrm{sw^{2}}}}\frac{u^{2}+2w^{2}}{u^{2}+w^{2}}\left[H_{2}^{+}\overleftrightarrow{\partial_{\mu}}H_{2}^{-}\right]Z^{\prime\mu}\label{eq:H2H2Zl}\end{eqnarray}

\end{document}